%% file: ms1.3apm.tex
\documentclass{aastex62} 
 
\usepackage{graphicx} 
\usepackage{txfonts} 
\usepackage{booktabs}  
\submitjournal{\apj}
\accepted{31/10/2018}
\shorttitle{Multiple populations in Galactic open clusters with Gaia} 
\shortauthors{G.\, Cordoni, et al.} 

\begin{document} 

\title{EXTENDED MAIN SEQUENCE TURNOFF AS A COMMON FEATURE OF MILKY WAY OPEN CLUSTERS}  
\author{G.\,Cordoni }
\affiliation{Dipartimento di Fisica e Astronomia ``Galileo Galilei'' -
  Univ. di Padova, Vicolo dell'Osservatorio 3, Padova, IT-35122}
\author{A.\ P.\,Milone}
\affiliation{Dipartimento di Fisica e Astronomia ``Galileo Galilei'' -
  Univ. di Padova, Vicolo dell'Osservatorio 3, Padova, IT-35122}
\author{A.\ F.\,Marino} 
\affiliation{Dipartimento di Fisica e Astronomia ``Galileo Galilei'' - Univ. di Padova, Vicolo dell'Osservatorio 3, Padova, IT-35122}
\affiliation{Research School of Astronomy \& Astrophysics, Australian National University, Canberra, ACT 2611, Australia}
\author{M.\, Di Criscienzo }
\affiliation{Istituto Nazionale di Astrofisica - Osservatorio Astronomico di Roma, Via Frascati 33, I-00040 Monteporzio Catone, Roma, Italy}
\author{F.\,D'Antona}
\affiliation{Istituto Nazionale di Astrofisica - Osservatorio Astronomico di Roma, Via Frascati 33, I-00040 Monteporzio Catone, Roma, Italy}
\author{A.\,Dotter}
\affiliation{Harvard-Smithsonian Center for Astrophysics, 60 Garden Street, Cambridge, MA 02138, USA}
\author{E.\ P.\,Lagioia }
\affiliation{Dipartimento di Fisica e Astronomia ``Galileo Galilei'' -
  Univ. di Padova, Vicolo dell'Osservatorio 3, Padova, IT-35122}
\author{M.\,Tailo}
\affiliation{Dipartimento di Fisica e Astronomia ``Galileo Galilei'' -
  Univ. di Padova, Vicolo dell'Osservatorio 3, Padova, IT-35122}   %
 
\correspondingauthor{G.\ Cordoni}
\email{giacomo.cordoni@phd.unipd.it} 
\begin{abstract}
We present photometric analysis of twelve Galactic open clusters and show that the same multiple-population phenomenon observed in Magellanic Clouds (MCs) is present in nearby open clusters.
  Nearly all the clusters younger than $\sim$2.5~Gyr of both MCs exhibit extended main-sequence turnoffs (eMSTOs) and all the cluster younger than $\sim$700~Myr show broadened/split main sequences (MSs). 
High-resolution spectroscopy has revealed that these clusters host stars with a large spread in the observed projected rotations.
  
In addition to rotation, internal age variation is indicated as a possible responsible for the eMSTOs, making these systems the possible young counterparts of globular clusters with multiple populations. Recent work has shown that the eMSTO+broadened MSs are not a peculiarity of MCs clusters. Similar photometric features have been discovered in a few Galactic open clusters, challenging the idea that the color-magnitude diagrams (CMDs) of these systems are similar to single isochrones and opening new windows to explore the eMSTO phenomenon. We exploit photometry+proper motions from Gaia DR2 to investigate the CMDs of open clusters younger than $\sim$1.5~Gyr. 

Our analysis suggests that: (i) twelve open clusters show eMSTOs and/or broadened MSs, that cannot be due neither to field contamination, nor binaries; (ii) split/broadened MSs are observed in clusters younger than $\sim$700~Myr, while older objects display only an eMSTO, similarly to MCs clusters; (iii) the eMSTO, if interpreted as a pure age spread, increases with age, following the relation observed in MCs clusters and demonstrating that rotation is the responsible for this phenomenon.
\end{abstract} 
 

\section{Introduction}\label{sec:intro}

In the past years, work based on high-precision {\it Hubble Space
  Telescope} ({\it HST}) photometry discovered that the 
color-magnitude diagrams (CMDs) of most star clusters younger than
$\sim$2.5~Gyr in the Large and Small Magellanic Cloud (LMC and SMC)
are not consistent with simple stellar populations. 
Specifically, most, if not all, of them exhibit extended main-sequence
turnoffs (eMSTOs, e.g.\,Mackey \& Broby Nielsen\,2007; Glatt et
al.\,2008; Milone et al.\,2009), and clusters
younger than $\sim$700~Myr display both eMSTOs and split main
sequences (MSs, e.g.\,Milone et al.\,2013; 2015; 2018; Li et al.\,2017; Correnti et al.\,2017). 
 
The comparison between the observed and synthetic CMDs from the Geneva
database (e.g.\,Georgy et al.\,2014) suggests that split MSs are consistent with 
two stellar populations with different rotation rates. A group of
stars with rotation close to the breakout value ($\omega \sim 0.9
\omega_{\rm cr}$), which corresponds to the red MS 
and includes about two thirds of the total number of MS stars, and a
population of slow rotators with $\omega \sim 0$, which populate the
blue MS 
(e.g.\,D'Antona et al.\,2015; Milone et al.\,2016). 
On the turn-off region, rapidly and slowly rotating stars distribute
on brighter and fainter magnitudes, respectively.
Measurements of rotational velocities in MS stars of the LMC cluster
 NGC\,1818 from high-resolution spectra collected with the Very Large
 Telescope (VLT) has recently provided direct evidence that the red-MS and the
 blue-MS stars exhibit different rotation rates (Marino et
 al.\,2018a). Similarly, high-resolution Magellan spectra confirm that
 the bright and the faint MSTO of NGC\,1866 are mostly populated by
 slow and fast rotators, respectively (Dupree et al.\,2017).     
 
 Although it is now widely accepted that rotation is one of the main
 driver for the 
photometric features appearing on the CMDs of young and
intermediate-age MC clusters, it might not be 
able to entirely reproduce the observations. 
Indeed, as noticed by Milone et al.\,(2017), a 
fraction of eMSTO are consistent with being younger than the bulk of cluster stars. 
It has been suggested that some clusters have experienced a prolonged
star formation, and that age variation, together with rotation is
responsible for the eMSTOs (e.g.\,Goudfrooij et al.\,2014; 2017). In
this case, the MC clusters could represent the younger counterparts of
the old globular clusters with multiple populations (e.g.\,Conroy et
al.\,2011; Keller et al.\,2011). As an alternative, D'Antona et
al.\,(2017) suggested that the evolution of
braked rapidly-rotating stars can mimic an age spread and contribute
to the eMSTO. 

 The recent discovery of eMSTOs in four open clusters, namely
 NGC\,2099, NGC\,2360, NGC\,2818, and NGC\,6705 has challenged the
 text-book concept that the CMDs of open clusters are proxy of single
 isochrone and have demonstrated that the eMSTO is not a peculiarity
 of Magellanic Cloud clusters (Marino et al.\,2018b). 
 Spectroscopy of MS stars in NGC\,6705 shows that the blue and the red
 MS are populated by slow and fast rotators, respectively (Marino et
 al.\,2018b). Similarly, the color and magnitude of eMSTO stars of
 NGC\,2818 and NGC\,6705 are connected with their rotational velocity
 (Bastian et al.\,2018; Marino et al.\,2018b). 
 These results suggest that rotation plays an important role in
 shaping eMSTOs and broadened or split MSs in Galactic open clusters,
 resembling Magellanic Cloud clusters. 
 
 In this work we exploit the Gaia data release 2 (DR2, Gaia
 collaboration et al.\,2018) to analyze photometry, parallaxes, and
 proper motions of a large sample of Galactic open clusters younger
 than $\sim$2~Gyr to investigate the occurrence of the eMSTO in their
 CMDs. 
 The paper is organized as follows. In Section~\ref{sec:data} we
 describe the dataset and the data analysis. The CMDs and the
 investigation of the presence of eMSTO and broadened MSs are discussed in
 Section~\ref{sec:cmds}. Section~\ref{sec:theory} presents a
 comparison of the data with theoretical models; while
 Section~\ref{sec:summary} is a summary and brief discussion of our results.

 \section{Data and data analysis} \label{sec:data}

To unambiguously identify multiple populations along the CMD, if present, we need densely-populated clusters with low differential reddening and
negligible contamination from field stars.
To do this, we selected all the Galactic open clusters of the new
general (NGC), Index (IC), Melotte, and Collinder catalogs that, according to Dias et al.\,(2002), have E(B$-$V)$<$0.35 and host more than 400 cluster
members. Moreover, we restrict our analysis to 
clusters older than 2.5~Gyr, as
there is no evidence of eMSTO and split MS in Magellanic Cloud
clusters with similar ages. Our sample also includes NGC\,6705
(M\,11, E(B$-$V)=0.43), which exceeds our reddening constraint,  
because previous evidence of eMSTOs and broadened MS has been
reported for this cluster (Marino et al.\,2018b).

\begin{figure} 
  \centering
  \includegraphics[width=14cm,trim={0 4.5cm 0 4.5cm},clip]{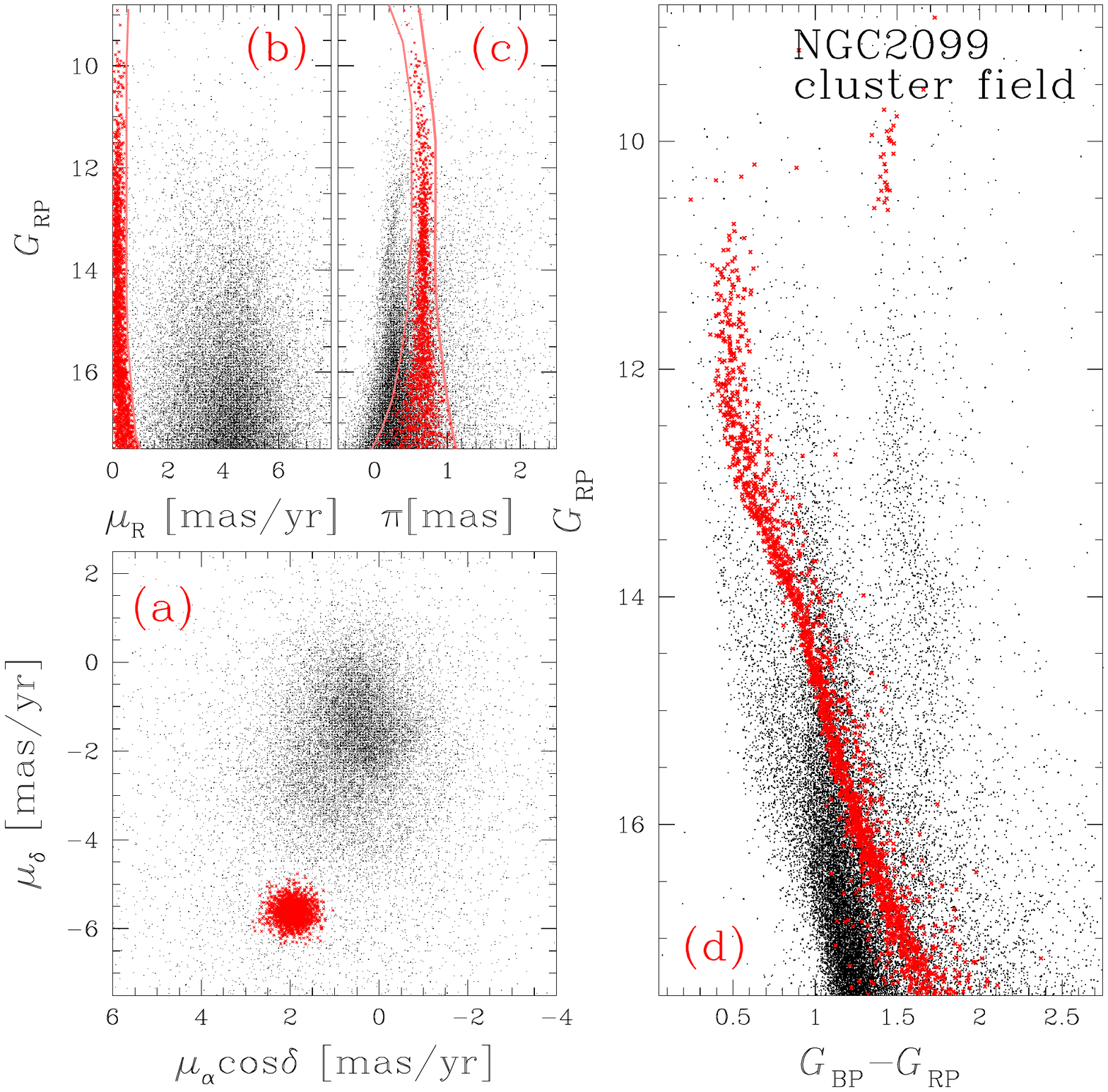}
  \caption{This figure illustrates the procedure that we used to
    select probable members of NGC\,2099. The VPD of
    proper motions for stars in the cluster field is plotted in panel
    a, while panels b and c show $G_{\rm RP}$ against proper motions
    and parallaxes, respectively. The red lines are used to separate
    NGC\,2099 members from field stars.  The $G_{\rm RP}$ vs.\,$G_{\rm
      BP}-G_{\rm RP}$ CMD is illustrated in panel d. Selected cluster
    members are represented with red symbols. See text for details.} 
 \label{fig:members} 
\end{figure}

We downloaded Gaia DR2 astrometry, photometry, parallaxes and proper motions
of stars within a radial distance from the center of
each cluster smaller than 2.5 times the cluster radius provided by
Dias et al.\,(2002) and 
identified a sample of cluster members by using the following iterative
procedure, which is illustrated in Fig.~\ref{fig:members} for
NGC\,2099.

\begin{itemize}
\item We first analyze the vector-point diagram (VPD) of stellar
  proper motions, and find that NGC\,2099 cluster members are clearly
  clustered around  ($\mu_{\rm \alpha}~cos{\delta}:\mu_{\rm \delta}$)$\sim$(1.9:$-$5.6). Hence we draw by eye a circle in the VPD
  that encloses most cluster members. 
The stars within the circle are selected to calculate the median
values of $\mu_{\rm \alpha}~cos{\delta}$ and $\mu_{\rm \delta}$
($<\mu_{\rm \alpha}~cos{\delta}>$ and $<\mu_{\rm \delta}>$) and to
derive the quantity $\mu_{\rm R}=\sqrt{ (\mu_{\rm
    \alpha}~cos{\delta}-<\mu_{\rm \alpha}~cos{\delta}>)^{2}   +
  (\mu_{\rm \delta}-<\mu_{\rm \delta}>)^2}$). 

\item  We plotted $G_{\rm RP}$ as a function of $\mu_{\rm R}$ for the
  selected stars and divided the analyzed magnitude interval with
  $10.0<G_{\rm RP}<17.5$ into bins of 0.5 mag each. For each bin we
  iterativelly calculated the median value of $\mu_{\rm R}$ ($\mu_{\rm
    R,  med}$) and the corresponding rms ($\sigma$) by rejecting all
  the stars with $\mu_{\rm R}>\mu_{\rm R,  med} + 4\cdot\sigma$. The
  mean magnitudes of each bin are associated to the quantities
  $\mu_{\rm R,  med} + 4\cdot\sigma$ and these points are linearly
  interpolated to derive the orange line plotted in the upper-left panel
  of Fig.~\ref{fig:members}. 
 Stars with deviations from $\mu_{\rm R, med}$ larger than $4\cdot\sigma$ are excluded from the sample of probable cluster members.

\item   We plotted $G_{\rm RP}$ as a function of the parallax,
  $\pi$, for the probable cluster members and calculated, for each
  bin of magnitude defined above, the median parallax $\pi_{\rm
    med}$ and the corresponding rms, $\sigma$, by using the same
  procedure described for proper motions. The orange lines plotted
  in the central-upper panel of Fig.~\ref{fig:members} are derived by
  adding $\pm 4 \cdot \sigma$ to $\pi_{\rm med}$ and all the stars
  that lie outside these two lines are excluded from the sample of
  probable cluster members. 

\item The selected stars are used to derive improved estimates of
  $<\mu_{\rm \alpha}~cos{\delta}>$ and $<\mu_{\rm \delta}>$. This ends
  one iteration. The procedure required three or four iterations to
  reach the convergence.  
\end{itemize}
  
\begin{figure}
  \centering
  \includegraphics[width=14cm,trim={0 5cm 0 4cm},clip]{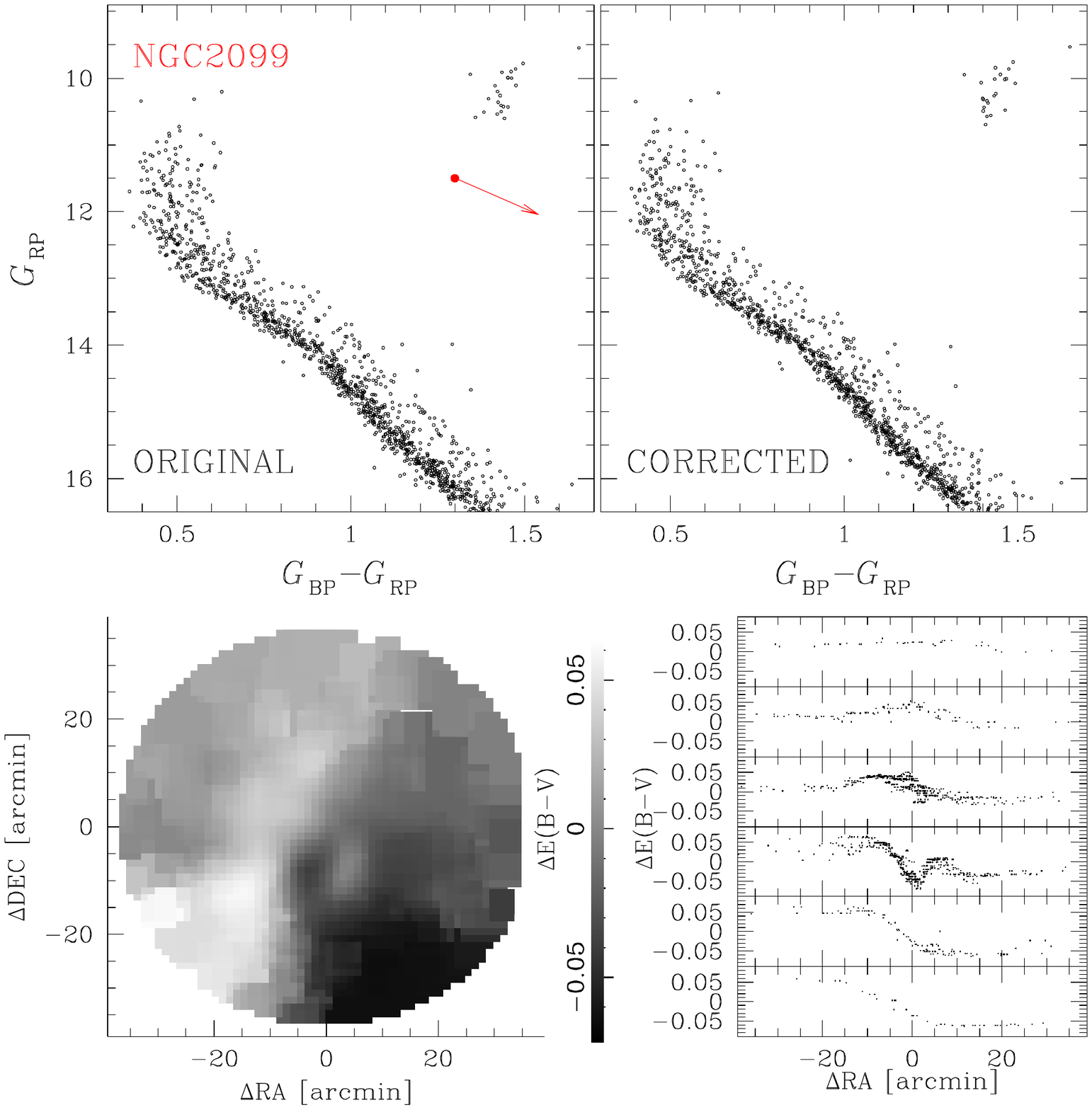}
  \caption{\textit{Upper panels.} Comparison of the original CMD of
selected cluster members of NGC\,2099 (left) with the CMD of the same
stars corrected for differential reddening (right). The arrow plotted
in the left-panel CMD indicates the reddening vector and corresponds
to $\Delta$E(B$-$V)=0.3 mag.
\textit{Lower panels.} Map of differential reddening, centered on
NGC\,2099. The levels of gray correspond to different E(B$-$V) values
as indicated by the scale on the middle (left). Right panels show
E(B$-$V) as a function of the right-ascension distance from the cluster
center for stars in six slices of declination.
   }
\label{fig:differential_reddening} 
\end{figure}

Photometry of cluster members has been corrected for differential
reddening by using the method described by Milone et al.\,(2012, see
their Sect.~1) and illustrated in Fig.~\ref{fig:differential_reddening}. 
Briefly, we first defined the reddening direction by using the
absorption coefficients in the $G_{\rm BP}$ and $G_{\rm RP}$ bands
provided by Casagrande \& VandenBerg\,(2018). Then, we derived the
fiducial of MS stars and calculated the color residuals from this
fiducial. To estimate the differential reddening suffered by each star
in the analyzed field of view, we selected a sample of 35 neighbours
formed by bright MS cluster members that are not evident binaries. Our
best differential-reddening estimate corresponds to the median of the
color residuals, calculated along the reddening line. To derive the
corresponding error, we subtracted the median from the residual of
each star and calculated the 68.27$^{\rm th}$ percentile of the
distribution of the corresponding absolute values ($\sigma$). We
considered the quantity 1.253$\cdot \sigma/\sqrt{35}$ as the
uncertainty associated to the differential-reddening.

As an example, we compare in Fig.~\ref{fig:differential_reddening} the original CMD of NGC\,2099 cluster members (upper-left panel) with the CMD corrected for differential reddening (upper-right panel). We also plot the differential-reddening map for a circular region with radius of 40 arcmin centered on NGC\,2099 (bottom left).  The bottom-right panel shows the reddening variation as a function of the  relative right-ascension distance from the cluster center for stars in six  declination intervals.

 A visual inspection of the differential-reddening corrected CMDs reveals that at least twelve open clusters, namely IC\,2714, Melotte\,71, NGC\,1245, NGC\,1817, NGC\,2099, NGC\,2360, NGC\,2818, NGC\,3114, NGC\,3532, NGC\,5822, and NGC\,6705, clearly exhibit multiple sequences in their CMDs. Their CMDs are presented and analyzed in the next section.

\section{Multiple populations along the color-magnitude diagrams}\label{sec:cmds}

\begin{figure*} 
  \centering
  \includegraphics[width=7.8cm,trim={0 4.5cm 0 4.5cm},clip]{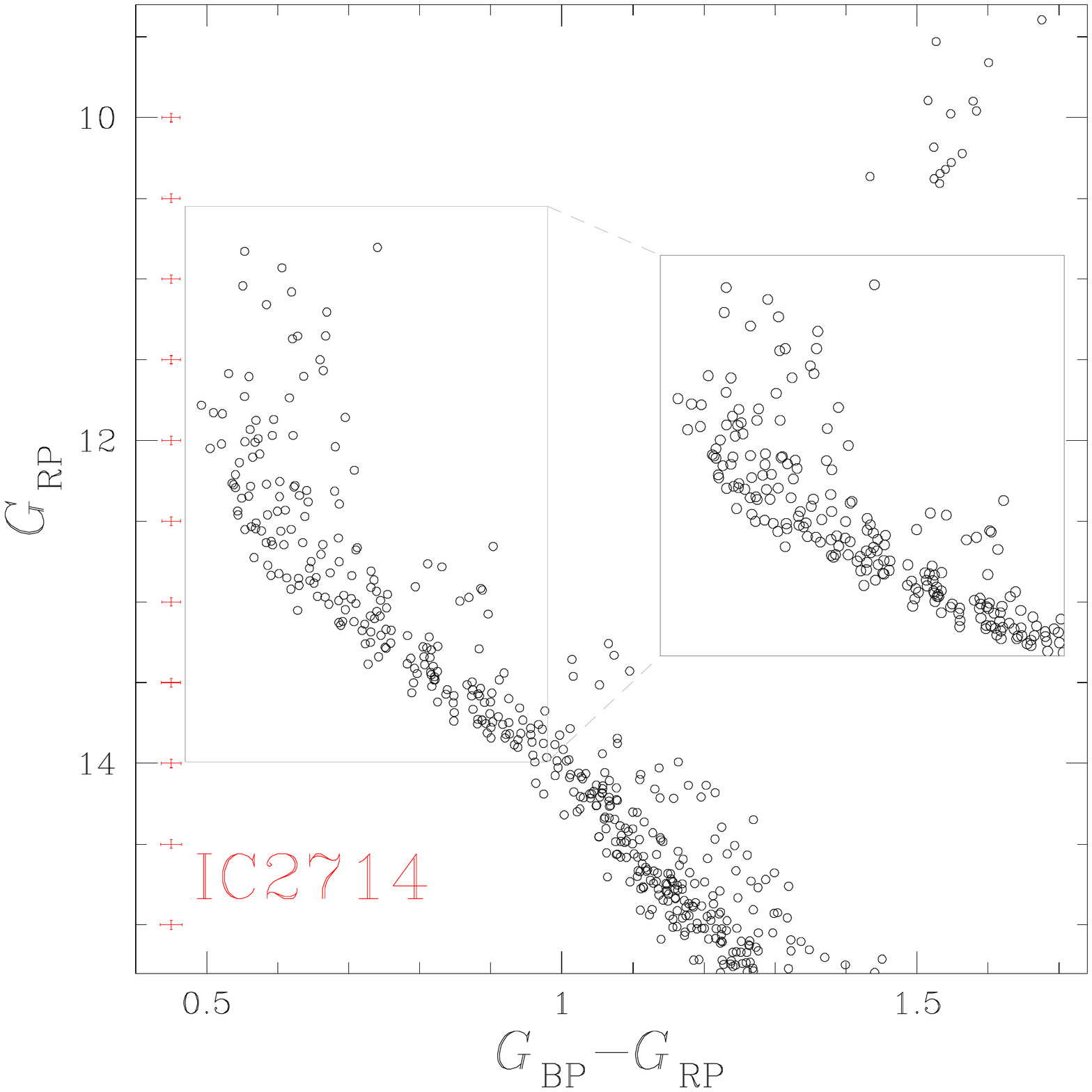}
  \includegraphics[width=7.8cm,trim={0 4.5cm 0 4.5cm},clip]{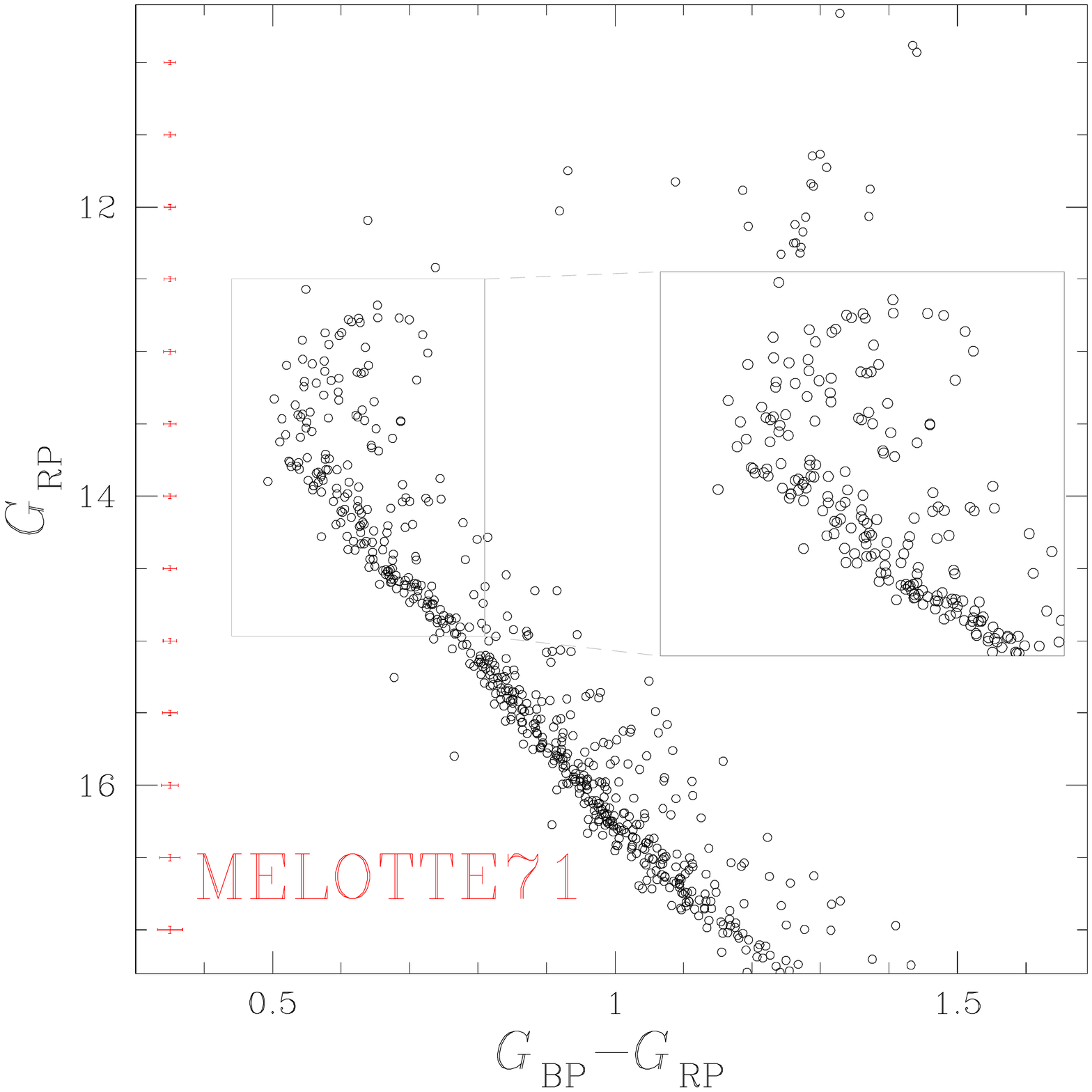}
  \includegraphics[width=7.8cm,trim={0 4.5cm 0 4.5cm},clip]{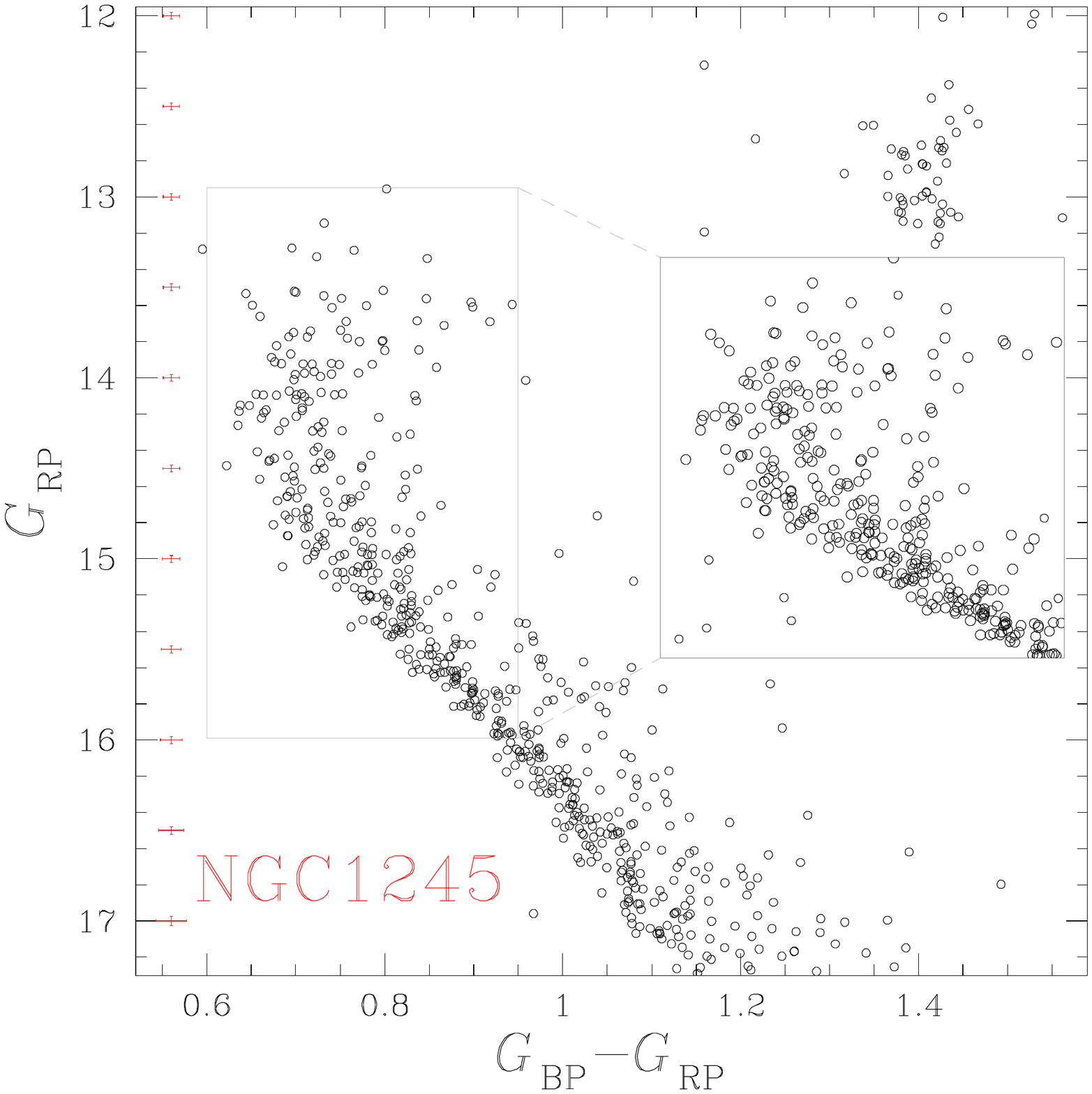}
  \includegraphics[width=7.8cm,trim={0 4.5cm 0 4.5cm},clip]{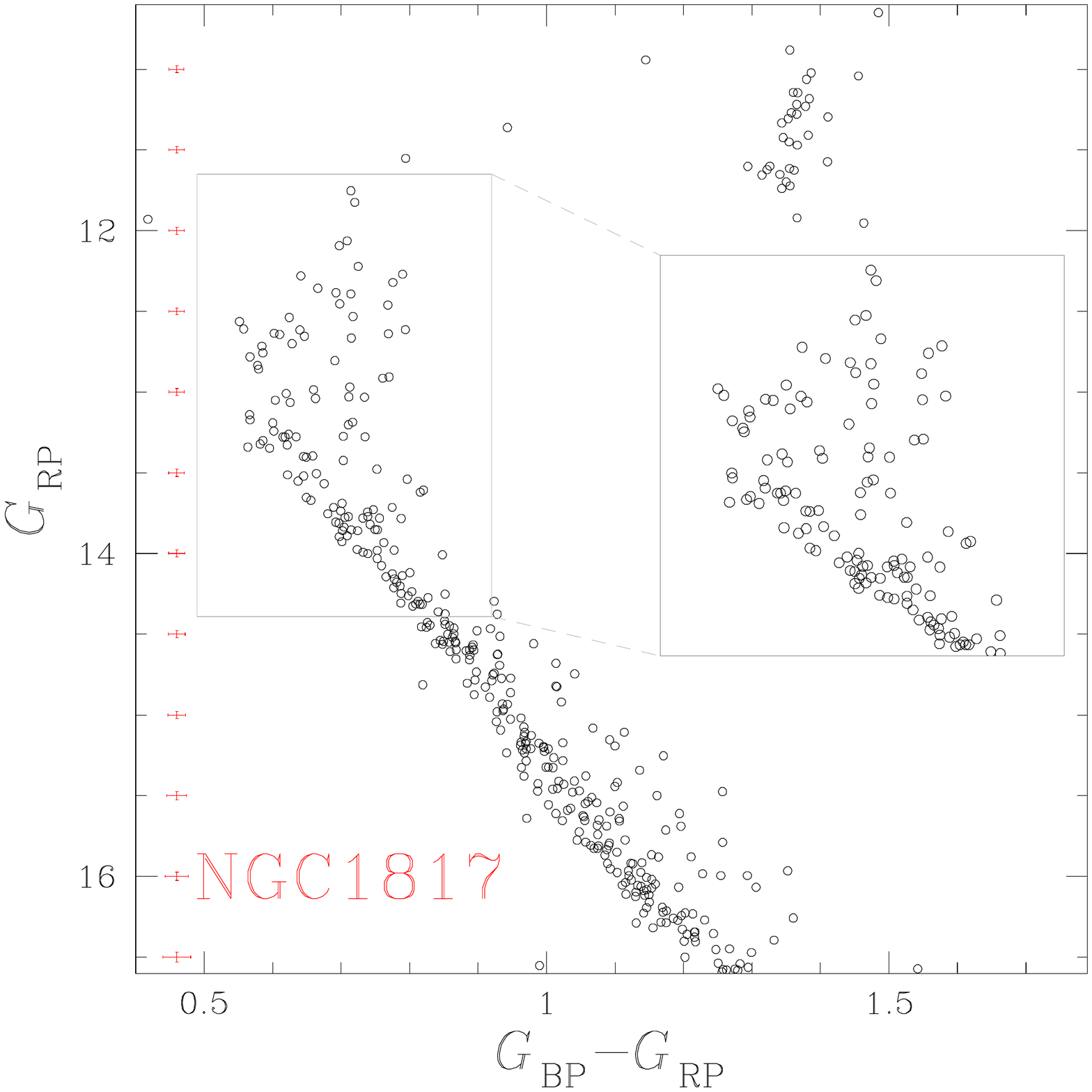}
  \includegraphics[width=7.8cm,trim={0 4.5cm 0 4.5cm},clip]{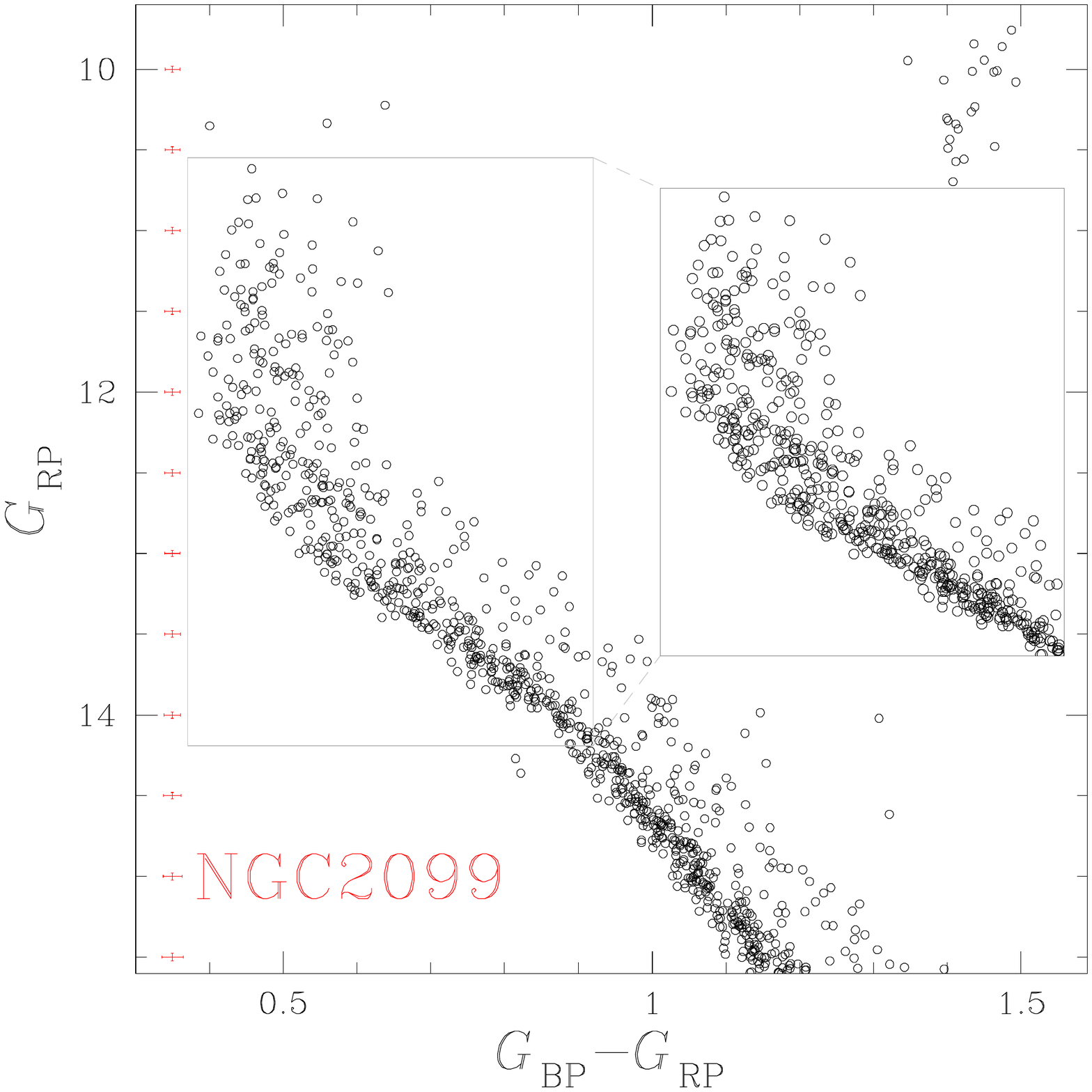}
  \includegraphics[width=7.8cm,trim={0 4.5cm 0 4.5cm},clip]{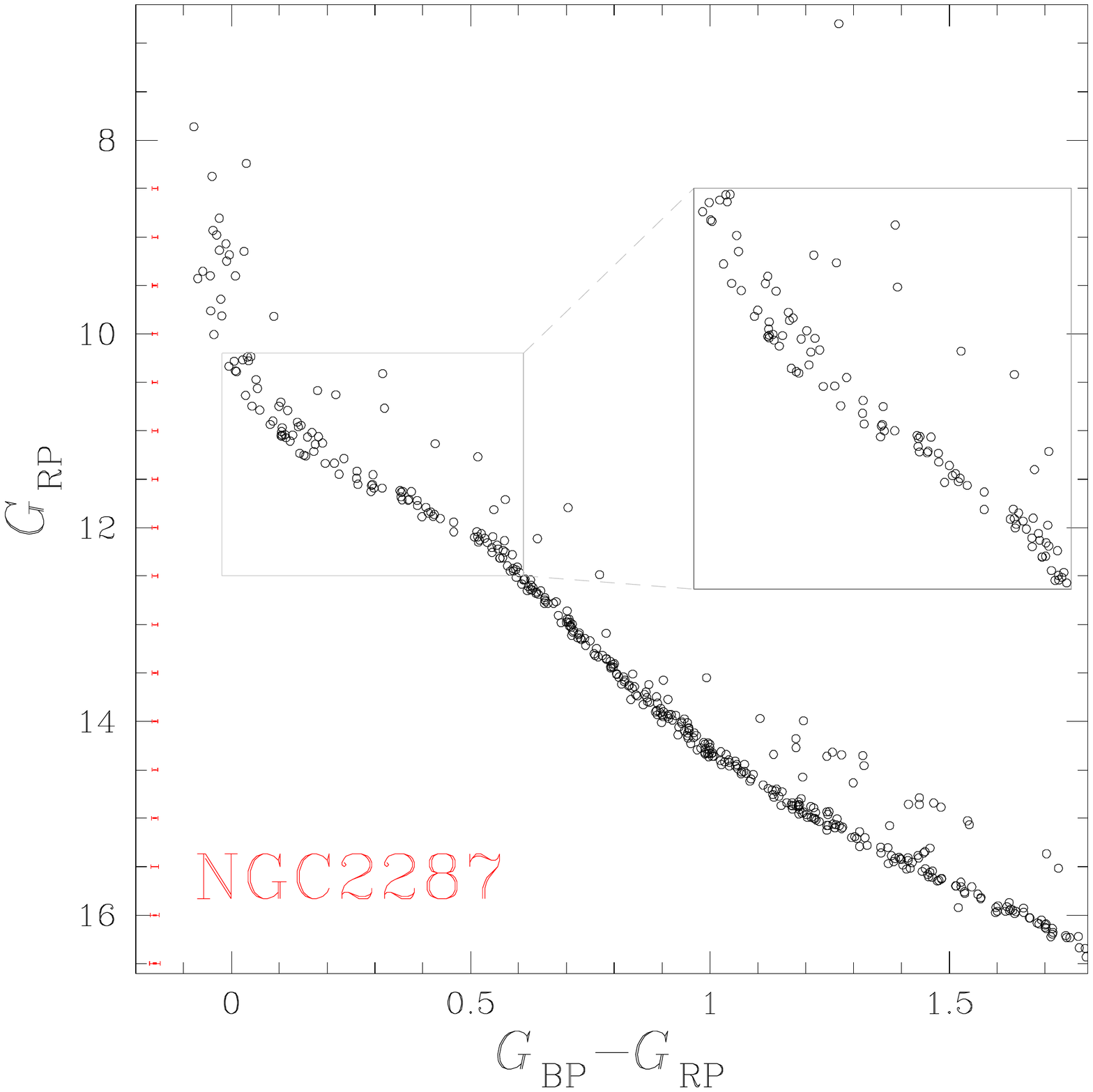}
  \caption{$G_{\rm RP}$ vs.\,$G_{\rm BP}-G_{\rm RP}$ CMDs, corrected
    for differential reddening, of cluster members for IC\,2714,
    Melotte\,71, NGC\,1245, NGC\,1817, NGC\,2099, and NGC\,2287. The
    insets highlight the eMSTO or the broadened MS. Red bars represent
  typical observational uncertainties.}
 \label{fig:CMDs1} 
\end{figure*}

\begin{figure*} 
  \centering
  \includegraphics[width=7.8cm,trim={0 4.5cm 0 4.5cm},clip]{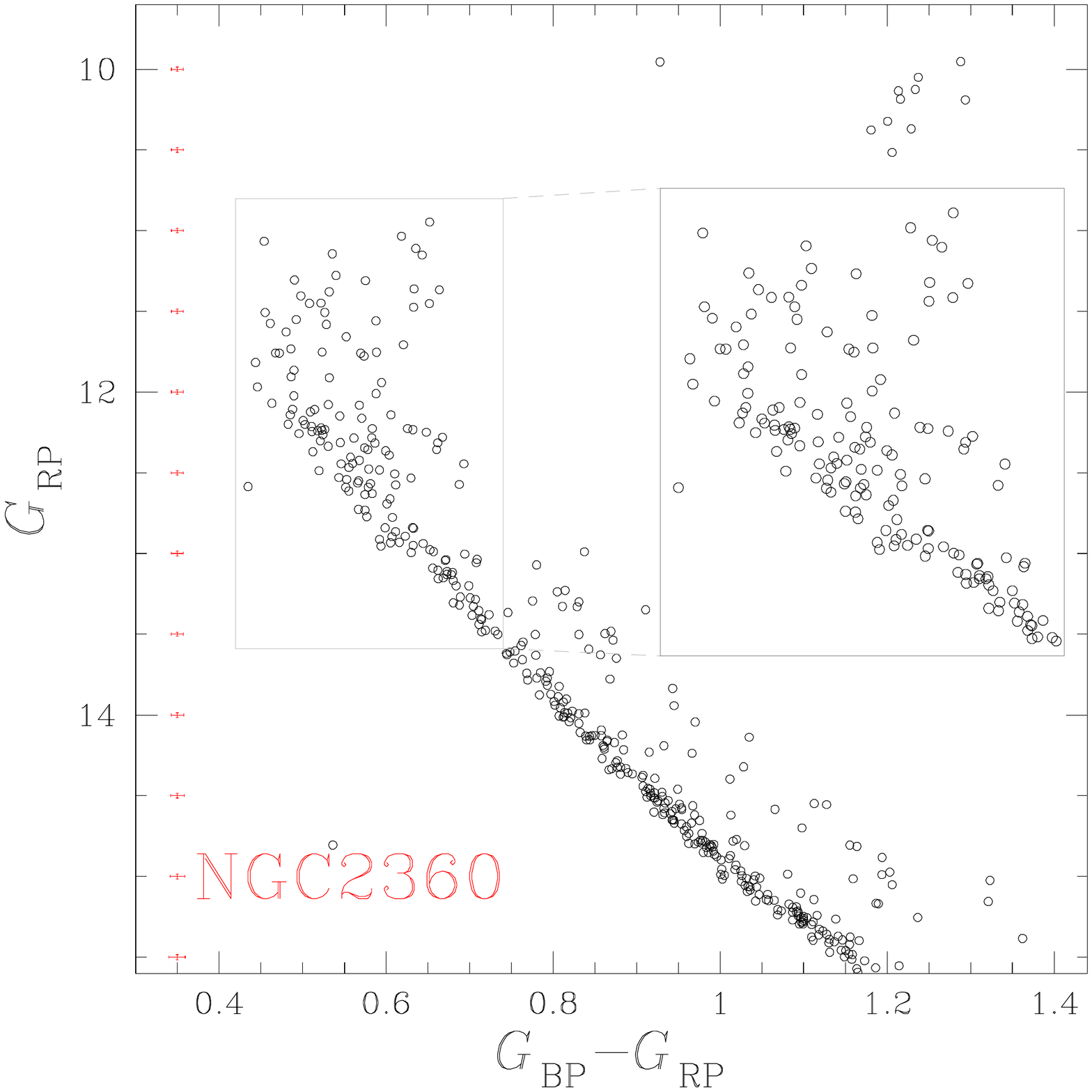}
  \includegraphics[width=7.8cm,trim={0 4.5cm 0 4.5cm},clip]{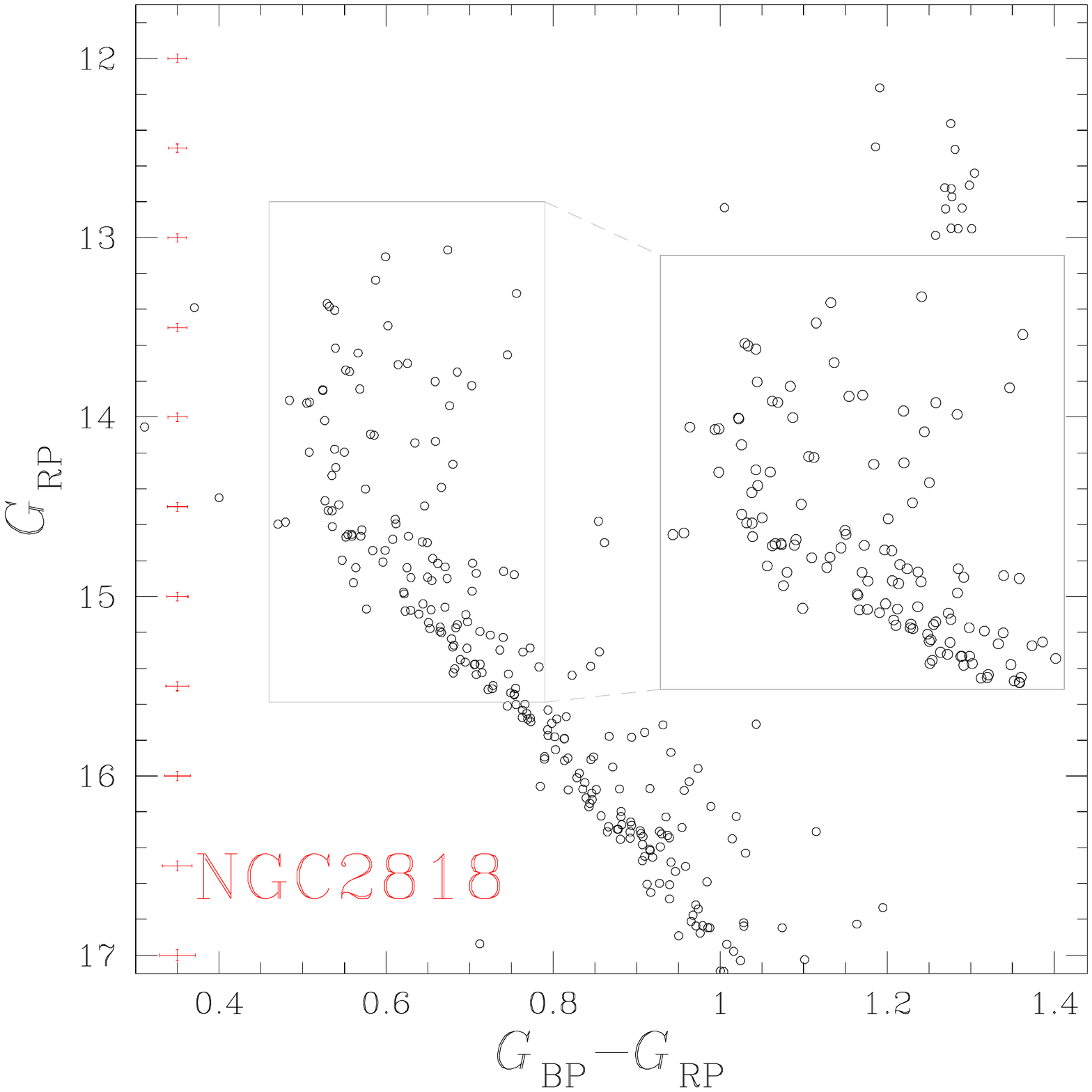}
  \includegraphics[width=7.8cm,trim={0 4.5cm 0 4.5cm},clip]{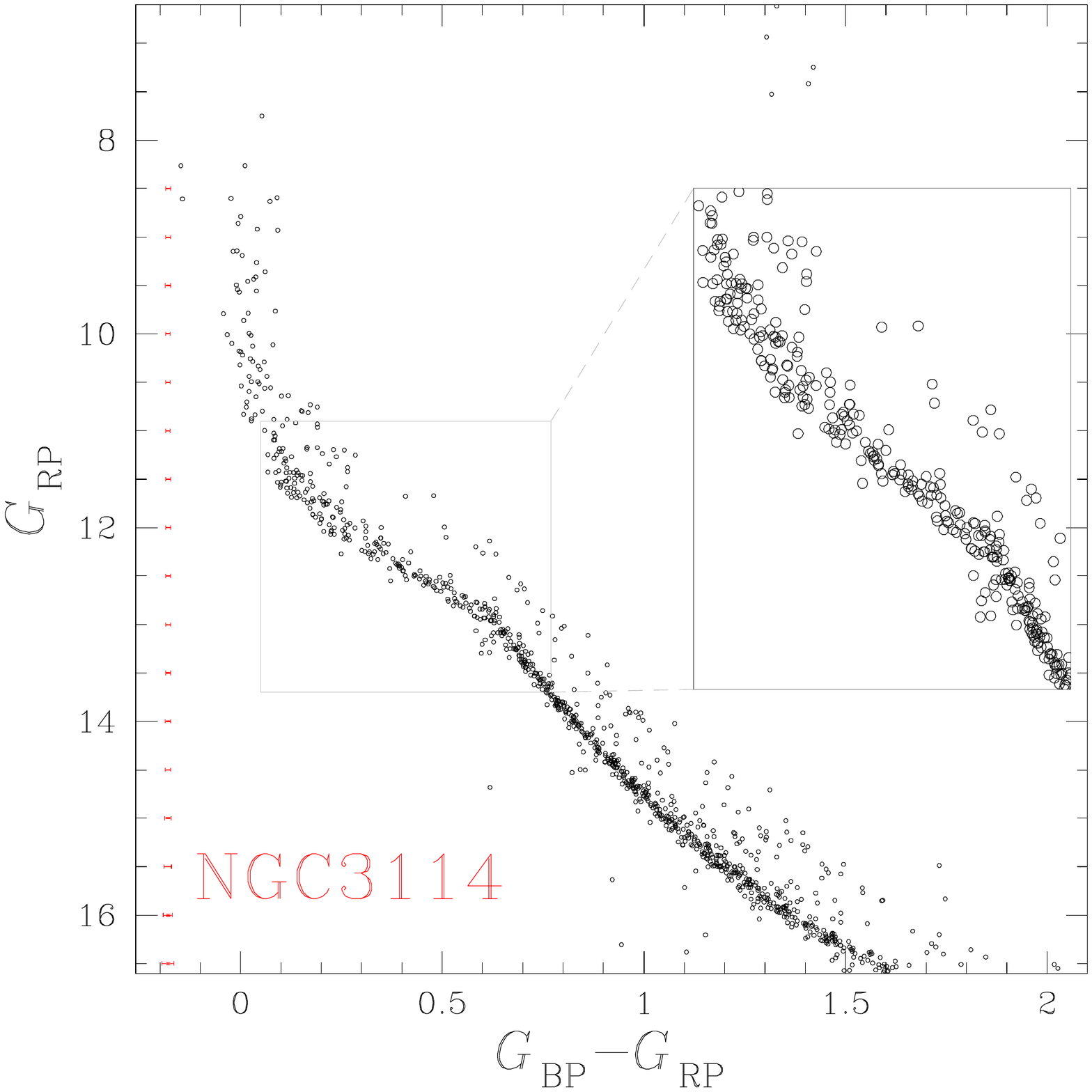}
  \includegraphics[width=7.8cm,trim={0 4.5cm 0 4.5cm},clip]{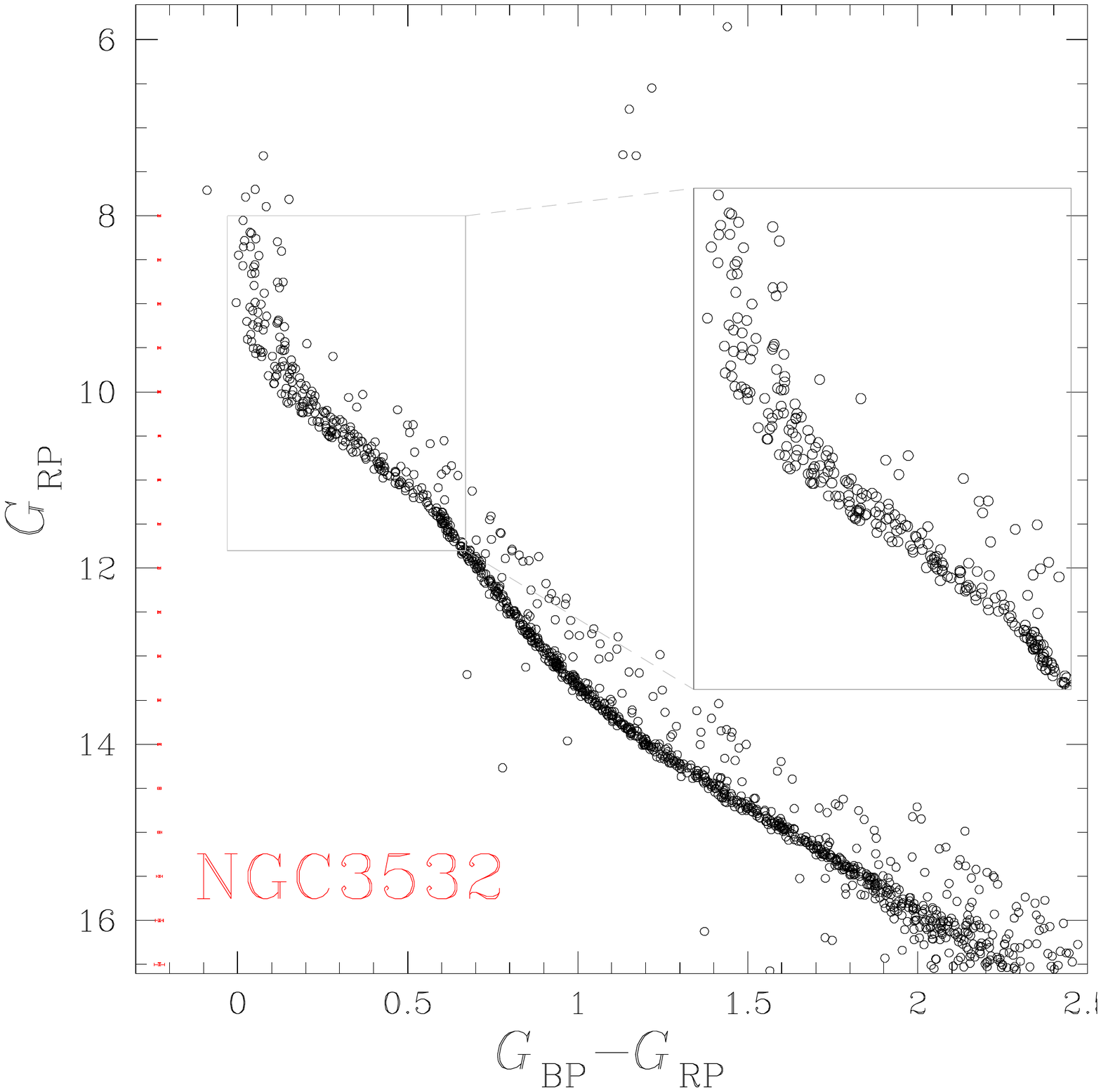}
  \includegraphics[width=7.8cm,trim={0 4.5cm 0 4.5cm},clip]{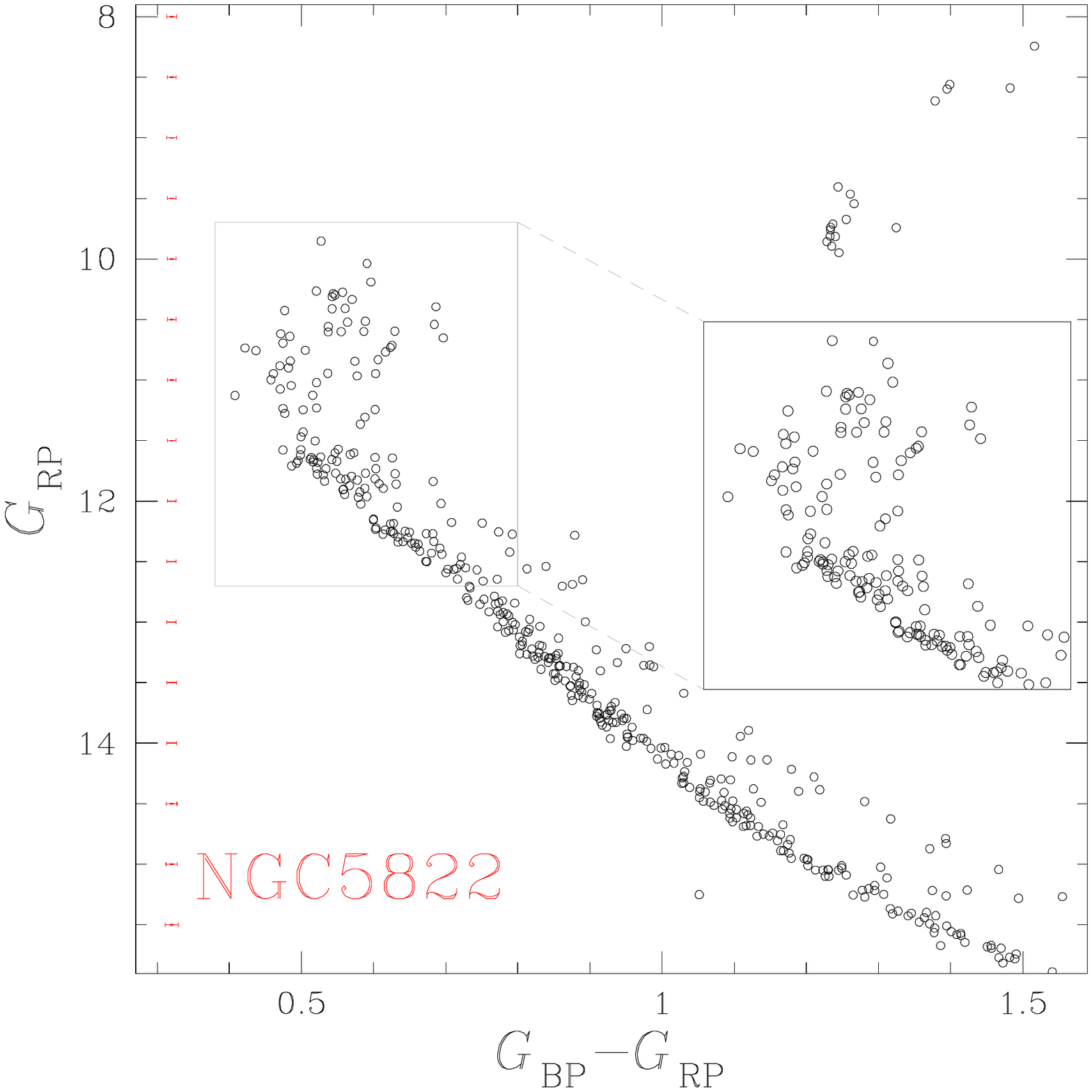}
  \includegraphics[width=7.8cm,trim={0 4.5cm 0 4.5cm},clip]{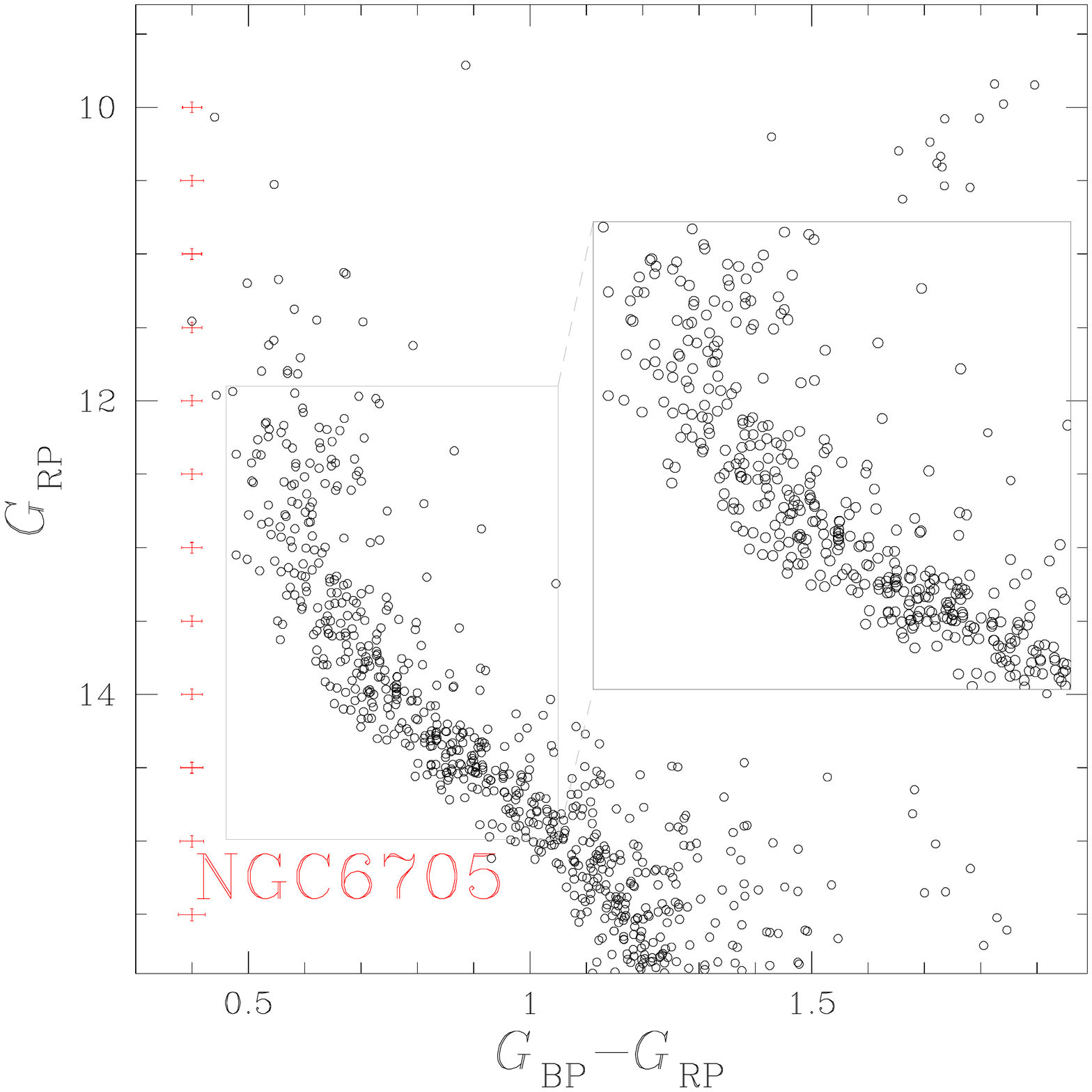}
  \caption{As in fig.~\ref{fig:CMDs1} but for NGC\,2360, NGC\,2818, NGC\,3114, NGC\,3532, NGC\,5822, and NGC\,6705.}
 \label{fig:CMDs2} 
\end{figure*}

The final CMDs, corrected for differential reddening, of the selected
cluster members are plotted in
Figs.~\ref{fig:CMDs1}--~\ref{fig:CMDs2}, where we also represent with
red error bars the typical observational uncertainties for stars with
different luminosities. A visual inspection of these figures clearly
reveals that IC\,2714, Melotte\,71, NGC\,1245, NGC\,1817, NGC\,2099,
NGC\,2360, NGC\,2818, NGC\,3114, NGC\,3532, NGC\,5822, and NGC\,6705
exhibit the eMSTO. Noticeably, the upper MS of NGC\,2099, NGC\,2287
(M\,41), NGC\,3114, NGC\,3532, and NGC\,6705 is broadened, in
contrast with the faint MS, which is narrow and well defined.
 Similarly to what previously observed in MCs clusters, the broadened
 MS seems to disappear at the luminosity of 
 the MS kink at $T_{\rm eff}\sim 7000$~K, which is a feature of the
 CMDs that indicates the onset on envelope convection due to the
 lowering of the adiabatic gradient in the region of partial hydrogen
 ionization (e.g.\,D'Antona et al.\,2002). 

 The eMSTOs and the broadened bright MSs are highlighted in the insets of
 Figs.~\ref{fig:CMDs1}--~\ref{fig:CMDs2}. 
In the following we 
demonstrate that they are intrinsic features of the cluster CMDs.
To this aim, we investigate the impact of observational uncertainties, residual
field-stars contamination and binaries on the appearance of eMSTOs and
the broadened MSs on the CMDs.

 Specifically, in Sect.~\ref{subsec:subtraction} we describe the
 method used to statistically subtract field stars with cluster-like
 parallaxes and proper motions from the CMDs of candidate cluster
 members; in Sect.~\ref{subsec:binaries} we estimate the fraction of
 binaries in each cluster; and in Sect.~\ref{subsec:simulations} we
 compare the observations with simulated CMDs that account for both
 binaries and observational errors.  


\subsection{Field stars}
\label{subsec:subtraction}

\begin{figure*} 
  \centering
  \includegraphics[width=14cm,trim={0 4.5cm 0 4.5cm},clip]{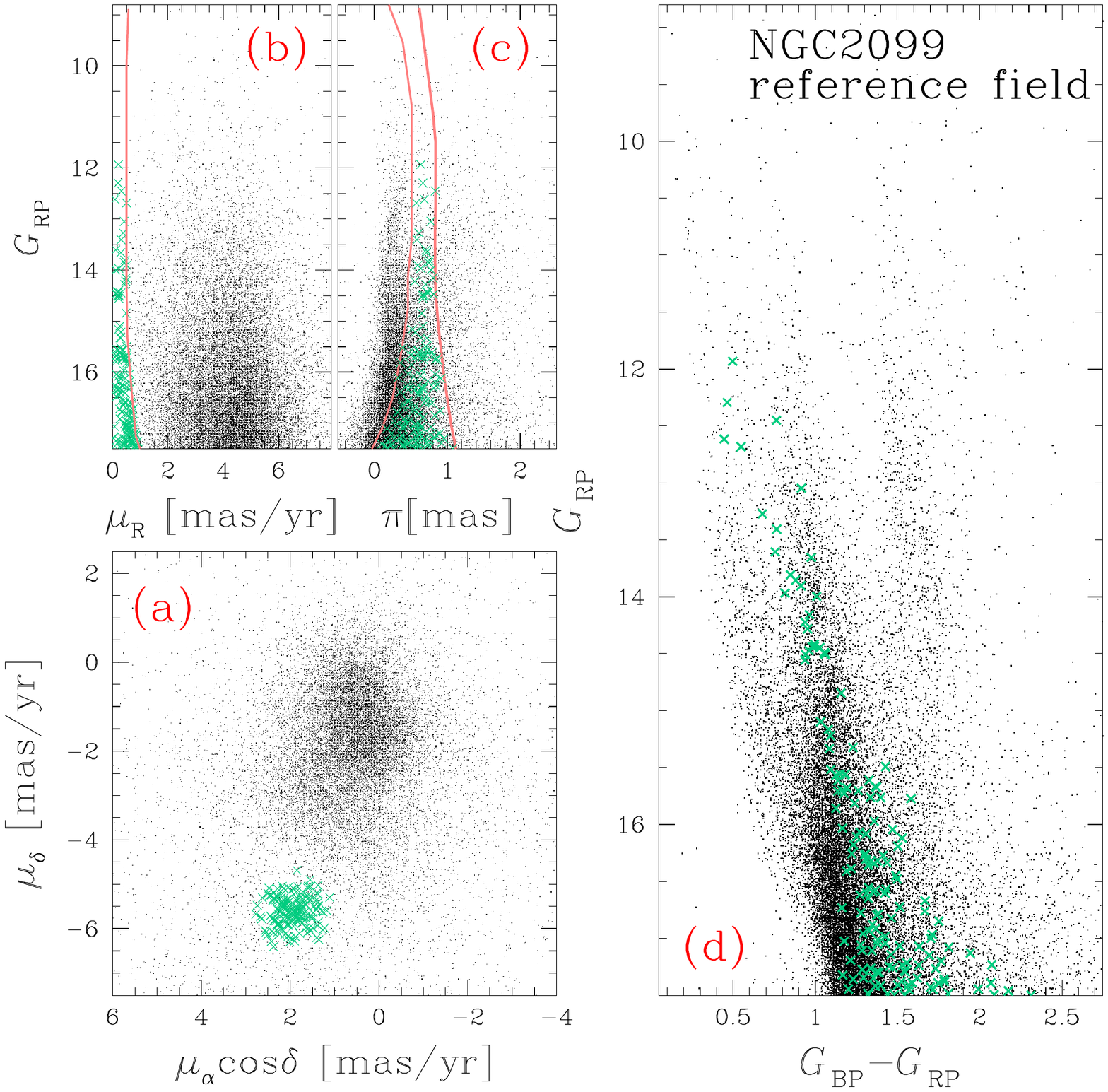}
  \caption{ This figure illustrates the procedure that we used to
    identify field stars with similar proper motions and parallaxes as
    NGC\,2099 cluster members. 
     The VPD of proper motions of stars in the ``reference field'' is plotted in panel a, while panels b and c show $G_{\rm RP}$ as a function of stellar proper motions and parallaxes, respectively. The orange lines defined in Fig.~\ref{fig:members} are overimposed to the diagrams of panels b and c.  The $G_{\rm RP}$ vs.\,$G_{\rm BP}-G_{\rm RP}$ CMD is illustrated in panel d. Selected field stars are marked with aqua crosses. See text for details.}
 \label{fig:residuals} 
\end{figure*}

The CMDs shown in Figs.~\ref{fig:CMDs1}--~\ref{fig:CMDs2} are mostly
populated by cluster members that have been selected on the basis of
their parallaxes and proper motions as described in
Sect.~\ref{sec:data}. 
To estimate the contamination from those field stars that have proper
motions and distances similar to those of cluster members, we applied
the procedure illustrated in Fig.~\ref{fig:residuals} for NGC\,2099. 

All the stars plotted in Fig.~\ref{fig:residuals} are located in the ``reference field'', which is a circular annulus with the same area as the cluster field, 
centered on the cluster and with internal radius corresponding to
three times the cluster radius provided by Dias et al.\,(2002). 
The VPD of proper motions for stars in the reference field is plotted in Fig.~\ref{fig:residuals}a;
panels b and c show the $G_{\rm RP}$ magnitude as a function of
parallax and proper motions, respectively. We plot in each panel the
orange lines derived in Fig.~\ref{fig:members} that are now used to select field stars with cluster-like proper motions and parallaxes, in close analogy with what we did for candidate cluster members. The stars with cluster-like proper motions have been selected according to their position in the diagrams plotted in  panels b and c and are represented with aqua crosses in all the panels of ~\ref{fig:residuals}.

To statistically subtract the selected field stars from the
cluster-field CMD we adopted 
the same procedure used in our previous papers (e.g.\,Milone et
al.\,2009). In a nutshell, we calculated for each selected star (i) in
the reference field a distance in the CMD 
$d_{\rm i} = \sqrt {  k ((G_{\rm BP, rf}^{\rm i}-G_{\rm RP, rf}^{\rm i})-(G_{\rm BP, cf}-G_{\rm RP, cf}))^{2} + (G_{\rm RP, rf}^{\rm i}-G_{\rm RP, cf})^{2}} $ \\
where $G_{\rm BP, rf (cf)}$ and $G_{\rm RP, rf (cf)}$ are the
magnitudes of the selected stars in the reference (cluster) field, and
$k=7$ is a constant that accounts for the fact that the color of a
star is better constrained than its magnitude (Gallard et al.\,2003;
Marino et al.\,2014). We finally subtracted the stars in the
cluster-field CMD with the smallest distance.

\subsection{Binaries}
\label{subsec:binaries}
Unresolved binaries formed by pairs of MS stars are redder and brighter than single MS stars with similar masses while binaries formed by a MSTO star and a MS or a MSTO star are brighter than the corresponding single MSTO stars.
In the following, we measure the fraction of MS-MS binary systems of each cluster to estimate the contribution of binaries to the eMSTO and the broadened MS.

\begin{figure*}
  \centering
  \includegraphics[width=14cm,trim={0 2cm 0 12cm},clip]{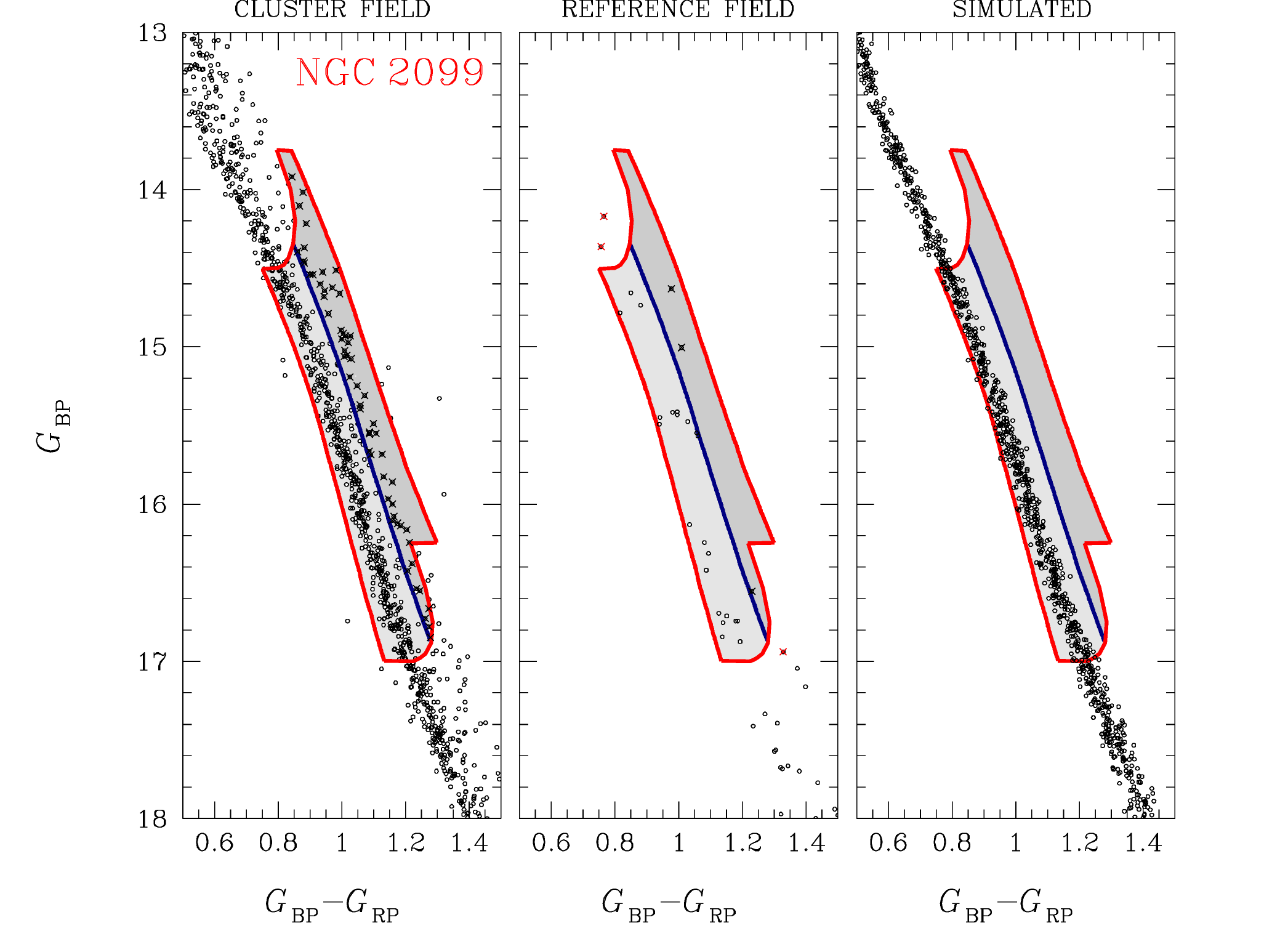}
  \caption{$G_{\rm BP}$ vs.\,$G_{\rm BP}-G_{\rm RP}$ CMD of selected NGC\,2099 cluster members in the cluster field (left panel) and CMD of stars with cluster-like proper motions and parallaxes in the reference field (middle panel). Right panel shows the simulated CMD. The shaded areas indicate the region A of the CMD, which is populated single MS stars and by MS-MS binary pairs with a primary component in the mass interval between 1.06 and 1.63 solar masses. The blue lines represent the fiducial lines of binaries with mass ratio, $q=0.7$. The region B of the CMD, which is populated by binaries with $q \geq 0.7$ (black crosses), is  colored dark-gray. See text for details.}
 \label{fig:binaries} 
\end{figure*}

To estimate the fraction of unresolved binaries with q$>$0.7 we used the procedure illustrated in Fig.~\ref{fig:binaries} for NGC\,2099, which is based on the method  by Milone et al.\,(2012, 2016) to characterize binaries in Galactic GCs.
We first identified two points along the MS with magnitudes $G_{\rm RP}^{\rm bright}$ and $G_{\rm RP}^{\rm faint}$, that delimit the MS region where the high-mass binaries are clearly separated from the remaining MS stars and there is no evidence for broadened or split sequences.   
We then defined two regions in the CMD, namely A and B, that correspond to the gray shaded areas in the CMDs of Fig.~\ref{fig:binaries}:
 region A includes all the single stars with $G_{\rm RP}^{\rm
   bright}<G_{\rm RP}<G_{\rm RP}^{\rm faint}$ and all the binaries
 with a primary component in the same magnitude interval; region B is
 the sub-region of A that is populated by binaries with q$>$0.7 and is
 represented with dark-gray colors in Fig.~\ref{fig:binaries}. 
The reddest line plotted in Fig.~\ref{fig:binaries} is the fiducial of equal-mass binaries shifted by four times the observational error in color to the red and the bluest line is the MS fiducial line shifted by four times the error in color to the blue. The fiducial of binaries with q$=$0.7 is represented by the blue continuous line and is derived by using the mass-luminosity relation inferred from the best-fit isochrone from Marigo et al.\,(2017).
For each cluster, we assumed the metallicity provided by Paunzen et
al.\,(2010), while
the adopted values of ages, reddening and distance modulus are those
providing the best match between the data and the isochrones and are
listed in Table~\ref{tab:results}.

The fraction of binaries is calculated as
\begin{equation}
  f_{\rm bin}^{\rm q>0.7}=\frac{N_{\rm cl}^{\rm B}-N_{\rm fi}^{\rm B}}{N_{\rm cl}^{\rm A}-N_{\rm fi}^{\rm A}}-\frac{N_{\rm sim}^{\rm B}}{N_{\rm sim}^{\rm A}}
  \end{equation}
 where $N_{\rm cl}^{\rm A,(B)}$ is the number of cluster members in the region A (B) of the CMD, $N_{\rm fi}^{\rm A,(B)}$ and $N_{\rm sim}^{\rm A,(B)}$ are the corresponding numbers of field stars with cluster-like proper motions and parallaxes and the number of simulated stars, respectively.

The measured fraction of binaries with q$>0.7$, $f_{\rm bin}^{\rm q>0.7}$ is used to extrapolate the total fraction of binaries, $f_{\rm bin}^{\rm TOT}$.
Specifically, by assuming a flat mass-ratio distribution, as observed among binaries with q$>0.5$ of Galactic GCs (Milone et al.\,2012, 2016), we infer $f_{\rm bin}^{\rm TOT} \sim 3.3 f_{\rm bin}^{\rm q>0.7}$. The total fraction binaries is typically around 0.30 and ranges from $\sim$0.11 for NGC\,2287 to $\sim$0.51 for NGC\,6705 and is similar to that observed in LMC clusters with similar ages (Milone et al.\,2009, see their Table~2). We thus confirm previous findings that open clusters typically host larger binary fraction than Galactic Globular Clusters (e.g.\,Sollima et al.\,2010).
 

\subsection{Simulated CMDs}
\label{subsec:simulations}
  The obtained total binary fractions, listed in Table~\ref{tab:results}, are used to simulate the CMD of a simple stellar population with the same observational errors, age, metallicity, distance modulus and reddening as inferred from the observations.
  To do this, we first associated to each star in the observed CMD of cluster member a synthetic star with the same magnitude and the color of the fiducial line. We selected a fraction of single stars equal to $f_{\rm bin}^{\rm TOT}$ and estimated the mass $\mathcal{M}$ of each of them by using the mass-luminosity relation by Marigo et al.\,(2017). We associated to each selected star a secondary star with a mass $\mathcal{M}_{2}=q \cdot \mathcal{M}$ and derived its $G_{\rm RP}$ magnitude from the relations by Marigo and collaborators. The corresponding color has been inferred from the fiducial line. Finally, we summed up the $G_{\rm BP}$ and $G_{\rm RP}$ fluxes of the two components, derived the corresponding magnitudes replaced the original star in the CMD with this binary system, and summed up the observational errors to all the stars of the CMD. 

\begin{figure*} 
  \centering
  \includegraphics[trim={0 0.9cm 0 0.0cm},clip,height=5.1cm,width=17cm]{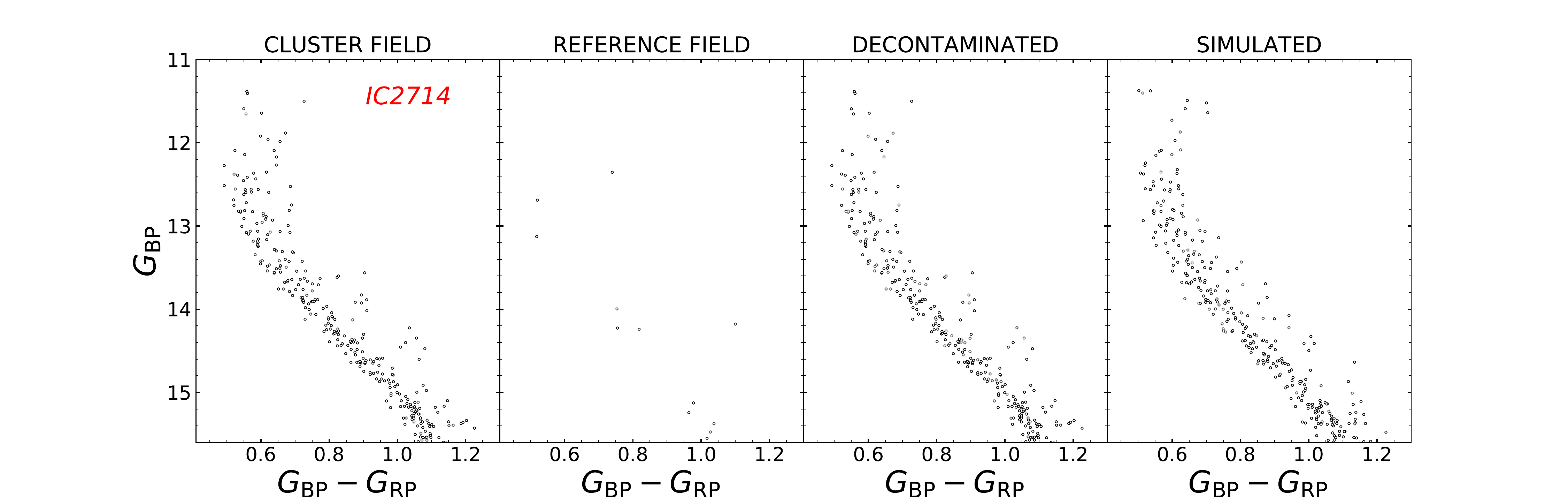}
  \includegraphics[trim={0 0.9cm 0 1.6cm},clip,height=4.5cm,width=17cm]{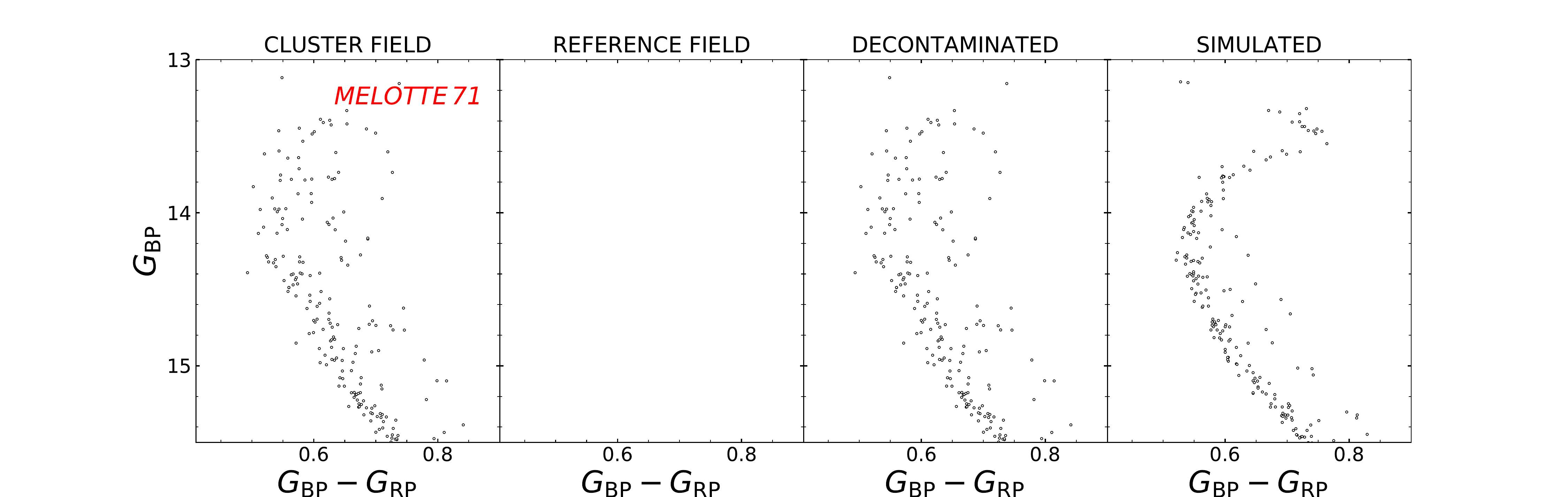}
  \includegraphics[trim={0 0.9cm 0 1.6cm},clip,height=4.5cm,width=17cm]{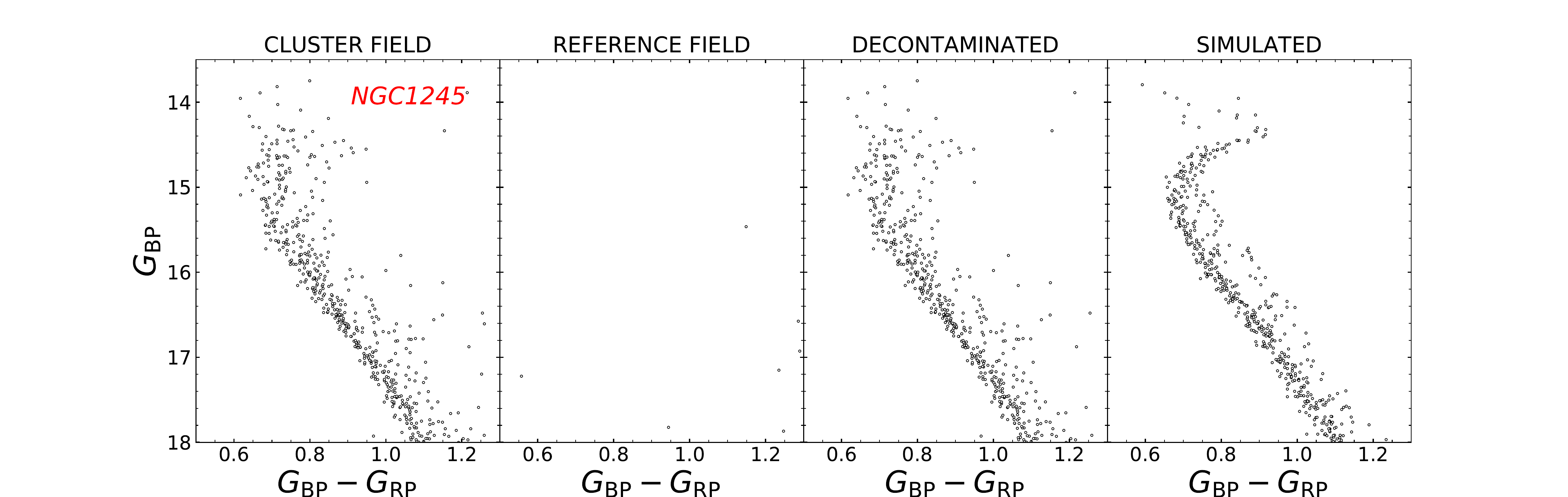}
  \includegraphics[trim={0 0.0cm 0 1.6cm},clip,height=5.0cm,width=17cm]{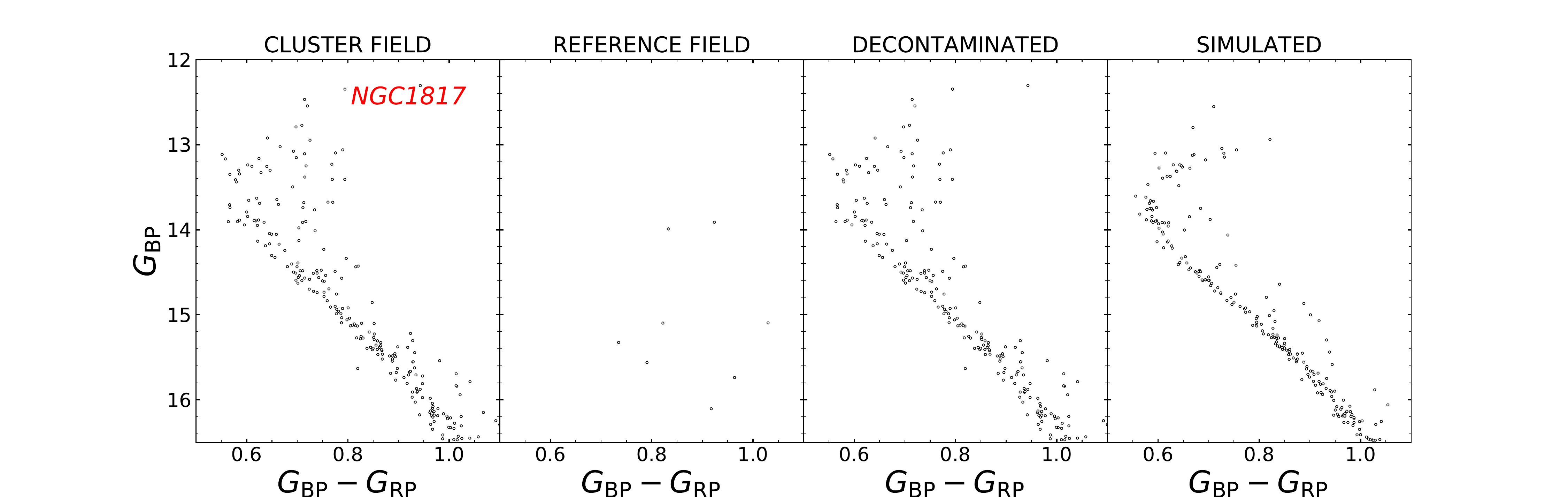}
  %
  \caption{From the left to the right. CMD of the selected cluster members in the cluster field for IC\,2714, Melotte\,71, NGC\,1245, and NGC\,1817 (first column), CMD of reference-field stars with cluster-like parallaxes and proper motions (second column), CMD of cluster members after the statistical subtraction of field stars with cluster-like parallaxes and proper motions (third column). Simulated CMD (forth column).   }
 \label{fig:ver1} 
\end{figure*}

\begin{figure*} 
\centering

\includegraphics[trim={0 0.9cm 0 0.0cm},clip,height=5.1cm,width=17cm]{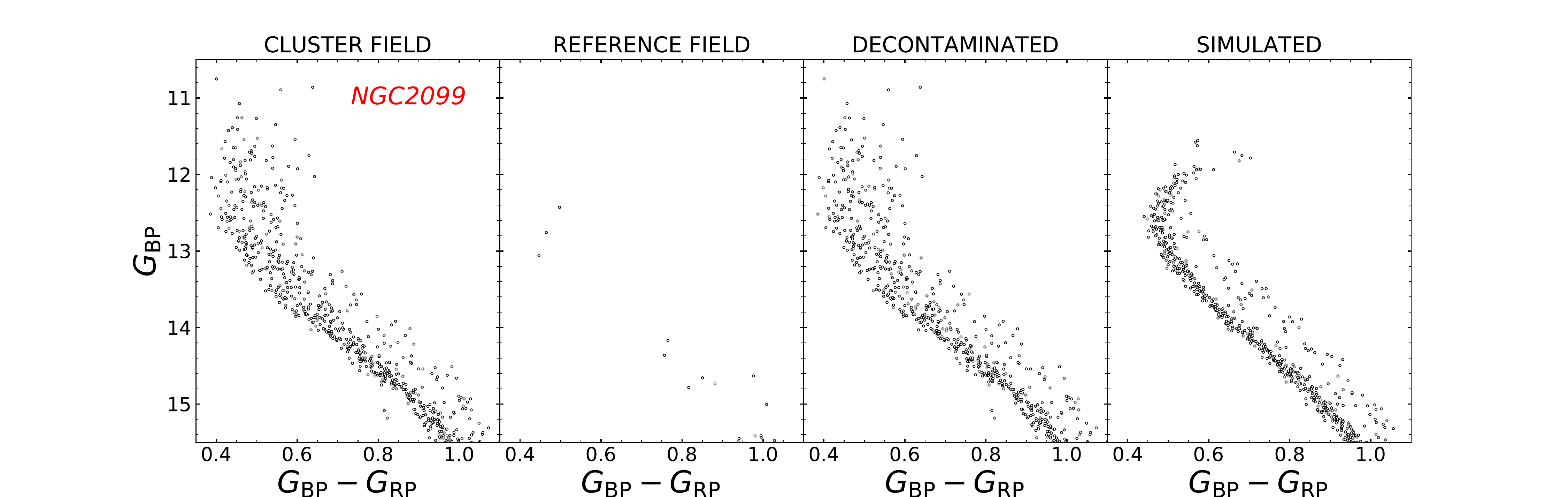}
\includegraphics[trim={0 0.9cm 0 1.6cm},clip,height=4.5cm,width=17cm]{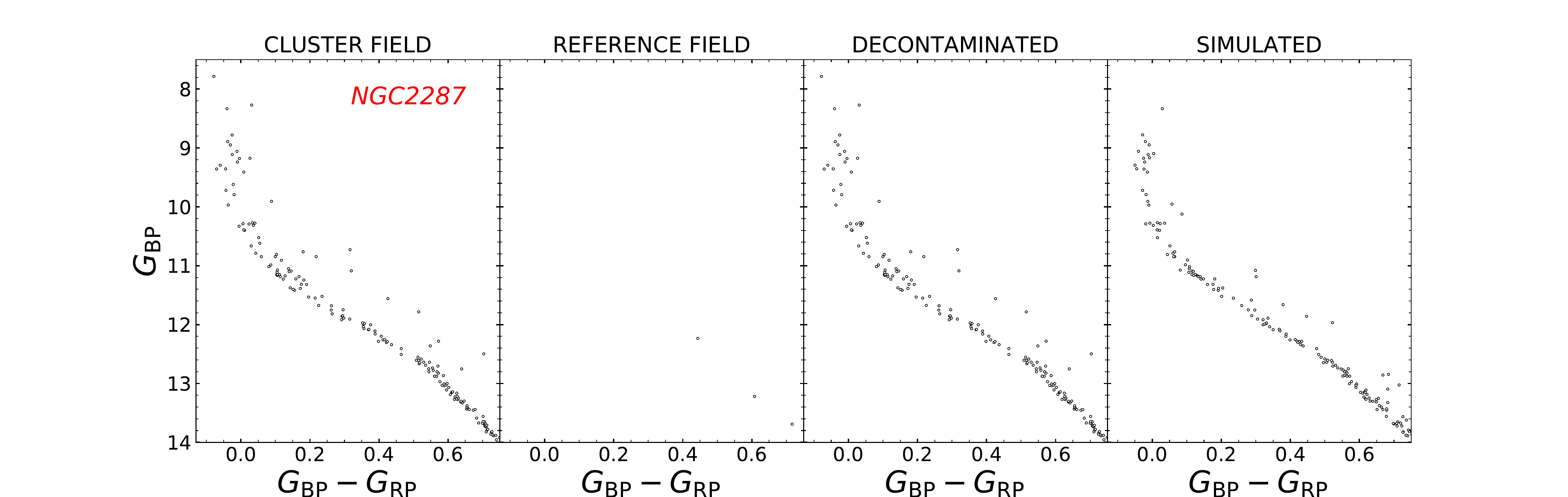}
\includegraphics[trim={0 0.9cm 0 1.6cm},clip,height=4.5cm,width=17cm]{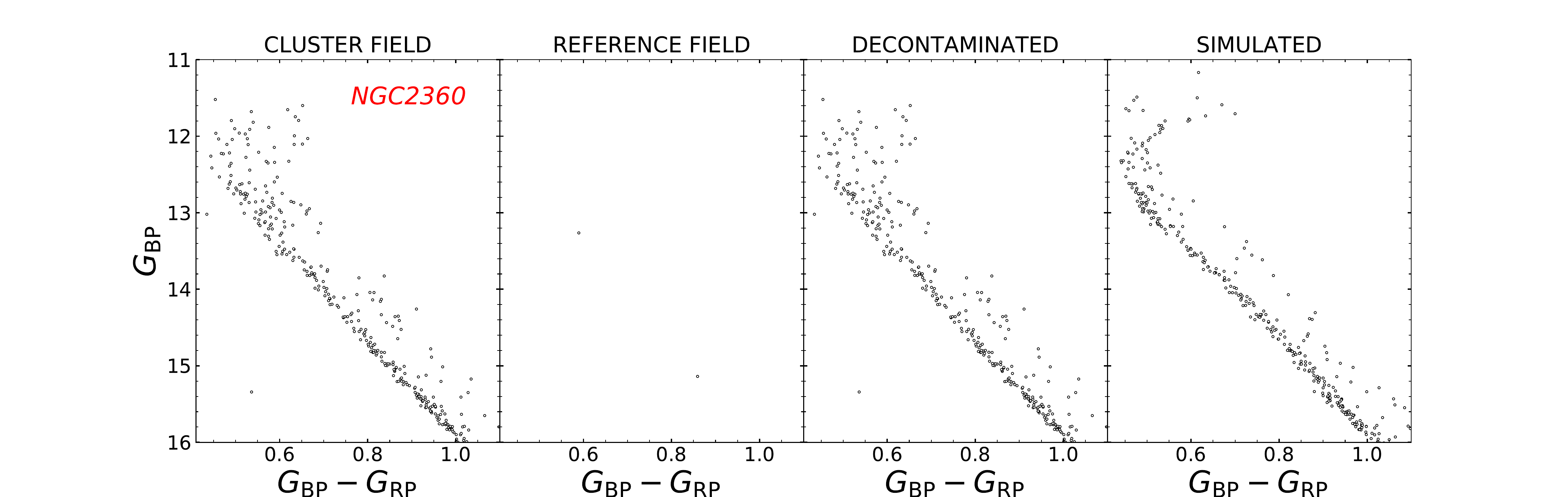}
\includegraphics[trim={0 0.0cm 0 1.6cm},clip,height=5.0cm,width=17cm]{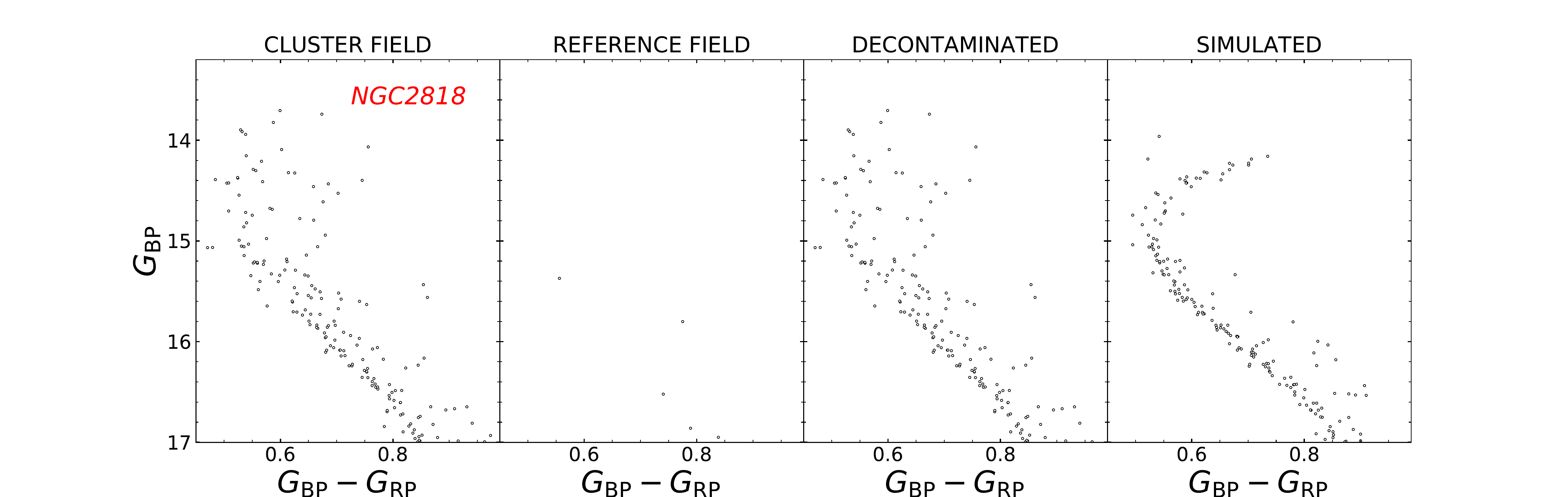}
  \caption{As in Fig.~\ref{fig:ver1} but for NGC\,2099, NGC\,2287, NGC\,2360 and NGC\,2818.}
 \label{fig:ver2} 
\end{figure*}

\begin{figure*} 
\centering
\includegraphics[trim={0 0.9cm 0 0.0cm},clip,height=5.1cm,width=17cm]{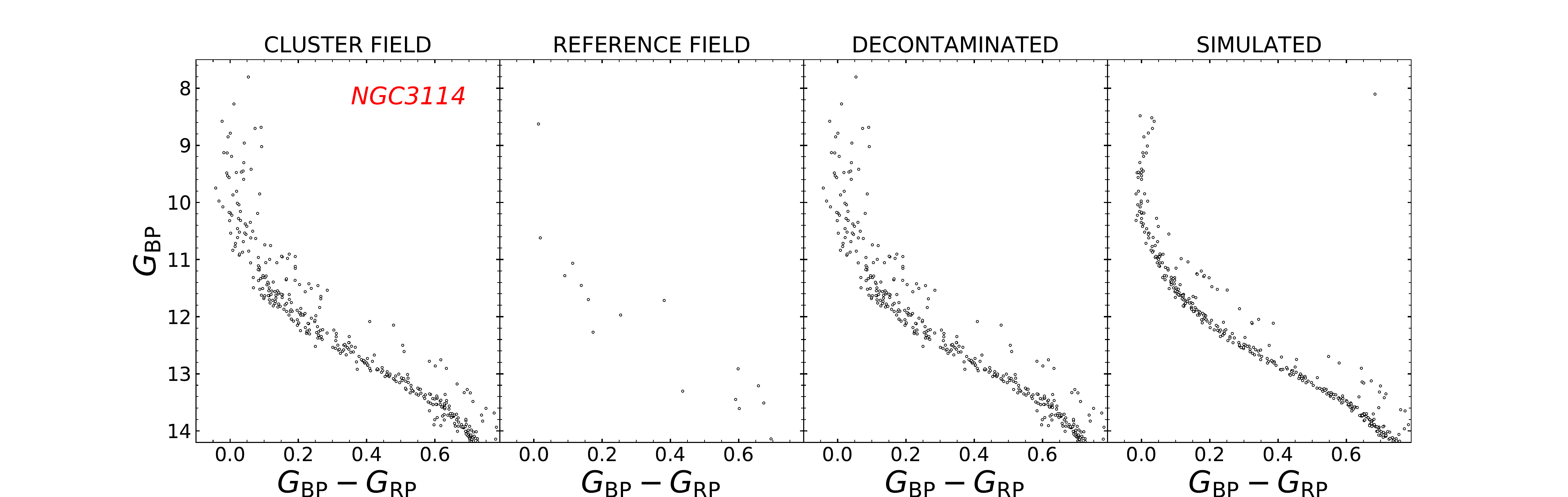}
\includegraphics[trim={0 0.9cm 0 1.6cm},clip,height=4.5cm,width=17cm]{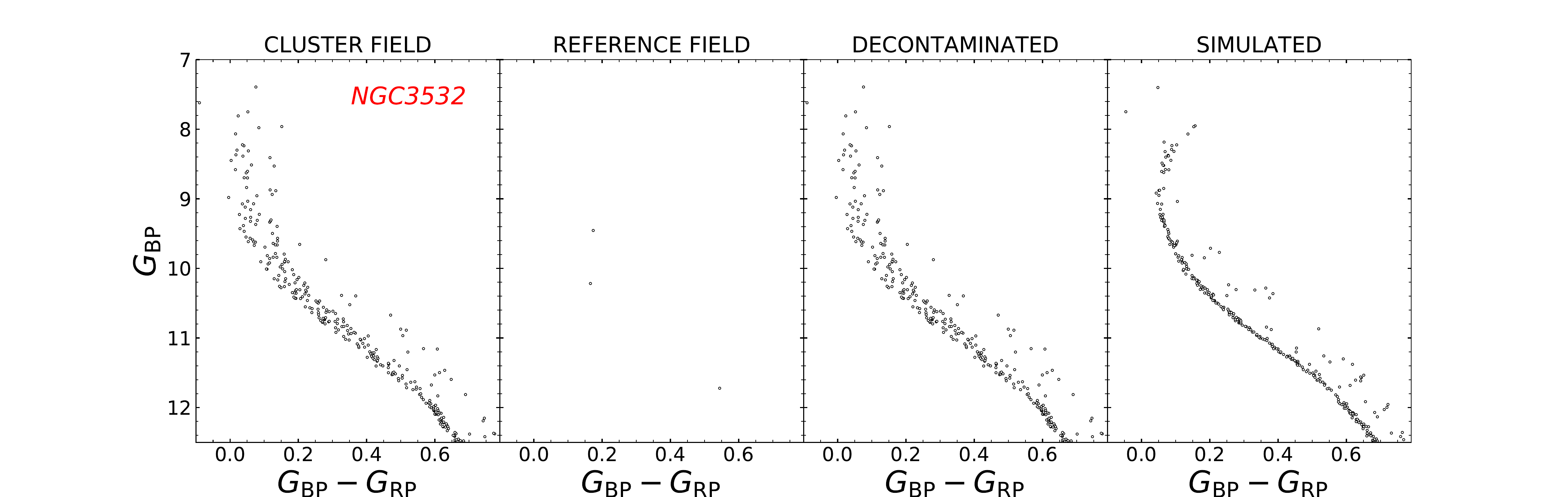}
\includegraphics[trim={0 0.9cm 0 1.6cm},clip,height=4.5cm,width=17cm]{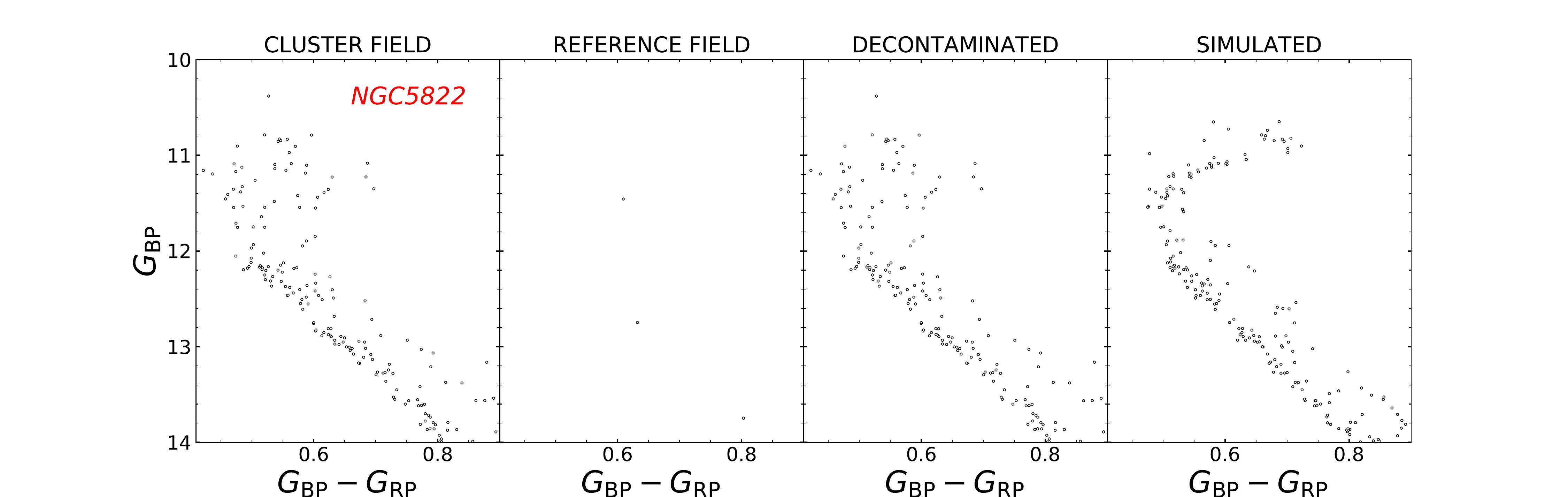}
\includegraphics[trim={0 0.0cm 0 1.6cm},clip,height=5.0cm,width=17cm]{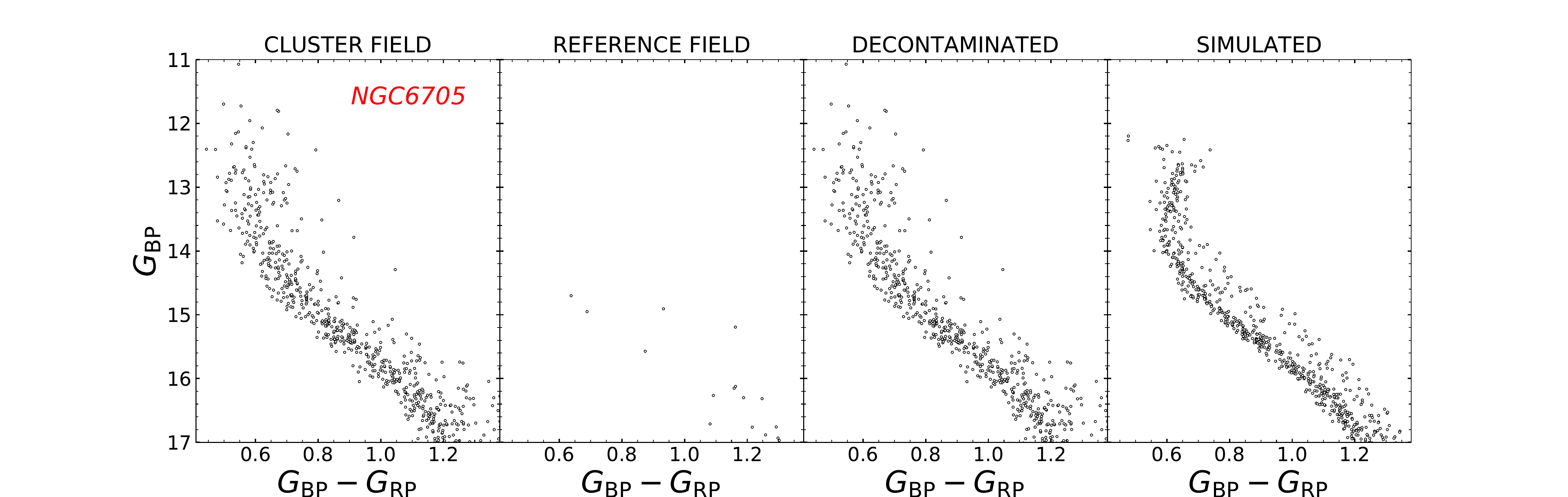}
  \caption{As in Fig.~\ref{fig:ver1} but for NGC\,3114, NGC\,3532, NGC\,5822 and NGC\,6705.}
 \label{fig:ver3} 
\end{figure*}

  Results are illustrated in Figs.~\ref{fig:ver1}--~\ref{fig:ver3}, where we compare for each cluster the CMD of selected cluster members (first column of panels), the CMD with cluster-like proper motions and parallaxes of stars in the reference field (second column), the decontaminated CMD (third column), and the simulated CMDs. 
 A visual inspection of these figures clearly demonstrates that the eMSTOs and the broadened MSs are not due neither to unresolved binaries nor to residual field-star contamination.

\section{Comparison with theory}\label{sec:theory}

The eMSTOs of Magellanic-Cloud clusters has been interpreted either as the signature of stellar populations with different ages (e.g.\,Mackey et al.\,2008; Goudfrooij et al.\,2011) or as the effect of stellar rotation on a single stellar population (e.g.\,Bastian \& De Mink 2009; Yang et al.\,2013; D'Antona et al.\,2016; Marino et al.\,2018). To disentangle between these two possibilities, in this section we compare the observed CMDs with isochrones of different ages and with simulated CMDs of stars with different rotation rates.

\begin{figure*} 
  \centering
  \includegraphics[width=15cm,trim={0 7cm 0 8cm},clip]{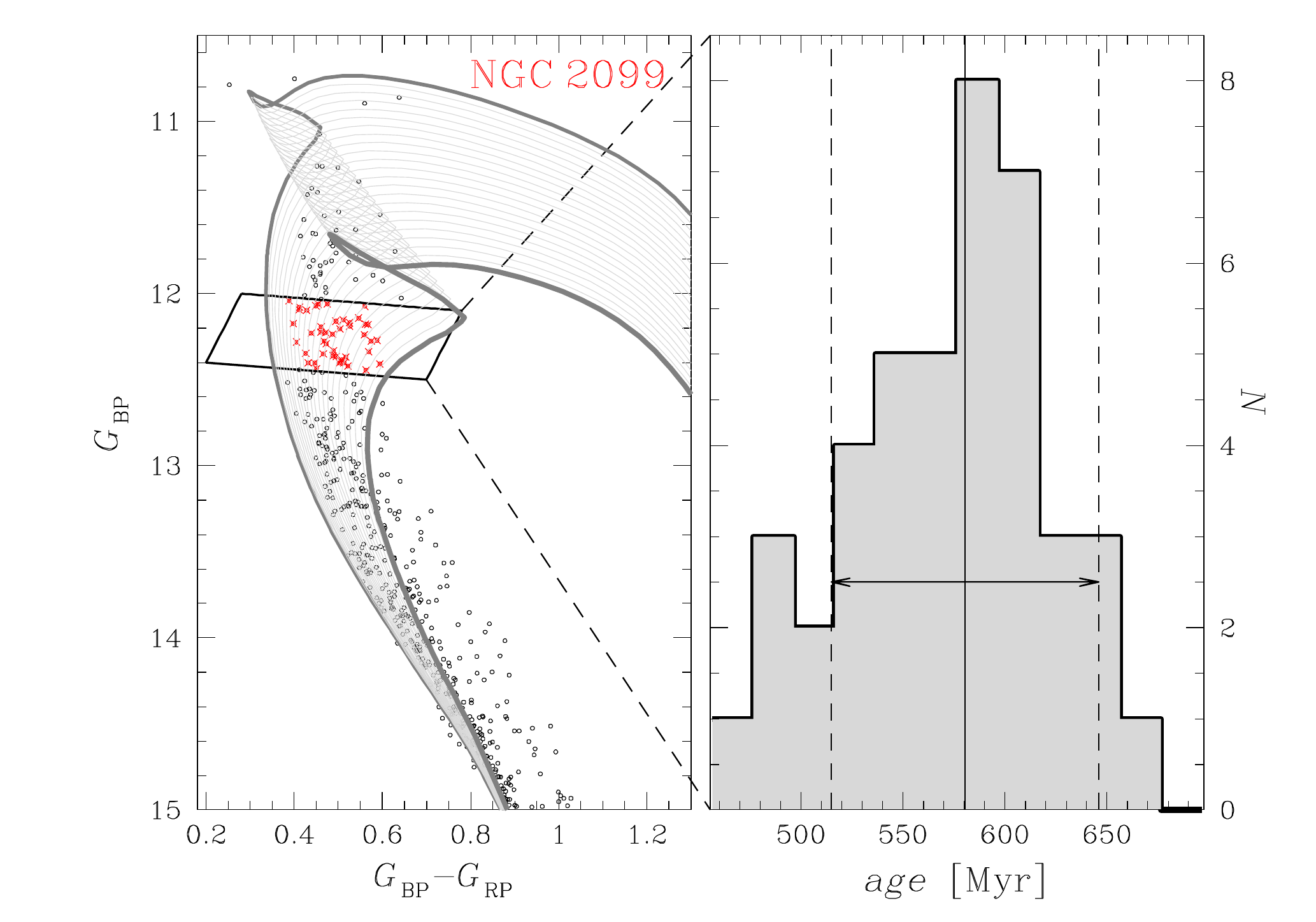}
  \caption{\textit{Left Panel.} Grid of isochrones from Marigo et al.\,(2017) overimposed on the CMD of NGC\,2099. The two isochrones represented with dark-tick lines have ages of 380 and 700 Myr, while the thin isochrones are spaced by 10 Myr in age. \textit{Right panel.} Histogram age distribution of the eMSTO stars plotted with red crosses in the left-panel CMD. The median age of these stars is marked with a vertical continuous line, while the two dashed lines have distances of $\pm \sigma$ from the median value. }
 \label{fig:agedis} 
\end{figure*}

To estimate the age spread, in the hypothesis that the eMSTO is
entirely due to a prolonged star formation, we compared the observed
CMDs with isochrones by Marigo et al.\,(2017). 
The procedure exploited to derive accurate age distributions is illustrated
in Fig.~\ref{fig:agedis} for NGC\,2099, and is similar to what we
have used in previous work (e.g.\,Milone et al.\,2015). 

In a nutshell, we first derived by hand the parallelepiped plotted in Fig.~\ref{fig:agedis} with the criterion of selecting the region around the turn off where the color and magnitude spread due to age variation are clearly distinguishable. Only stars within the parallelepiped are used to infer the age distribution.
Then, we overimposed on the CMD a grid of isochrones with the same metallicity and $[\alpha/Fe]$ and ages between 380 and 700~Myr in steps of 10~Myr (grey lines in Fig.~\ref{fig:agedis}) and derived isochrones separated by 1~Myr by linearly interpolating among these isochrones. We associated to each star the age of the closest isochrone and derived the age distribution shown in the right panel of Fig.~\ref{fig:agedis}. Finally, we calculated the median age and the absolute value of the difference between the age of each star and the median. We considered the 68.27$^{\rm th}$ percentile of the distribution of these absolute values as indicative of the observed age spread, $\sigma_{\rm AGE, obs}$. To estimate the contribution of observational errors on the inferred age spread, we applied the procedure described above to the simulated CMD of a simple population and derived the corresponding age spread,  $\sigma_{\rm AGE, sim}$. The intrinsic age spread is estimated as $\sigma_{\rm AGE}=\sqrt{\sigma_{\rm AGE, obs}^{2}-\sigma_{\rm AGE, sim}^{2}}$.
Uncertainties on $\sigma_{\rm AGE}$ are derived by bootstrapping with replacements performed 1,000 times on both the observed and the simulated age distributions. 

\begin{figure}
  \centering
  \includegraphics[width=13cm]{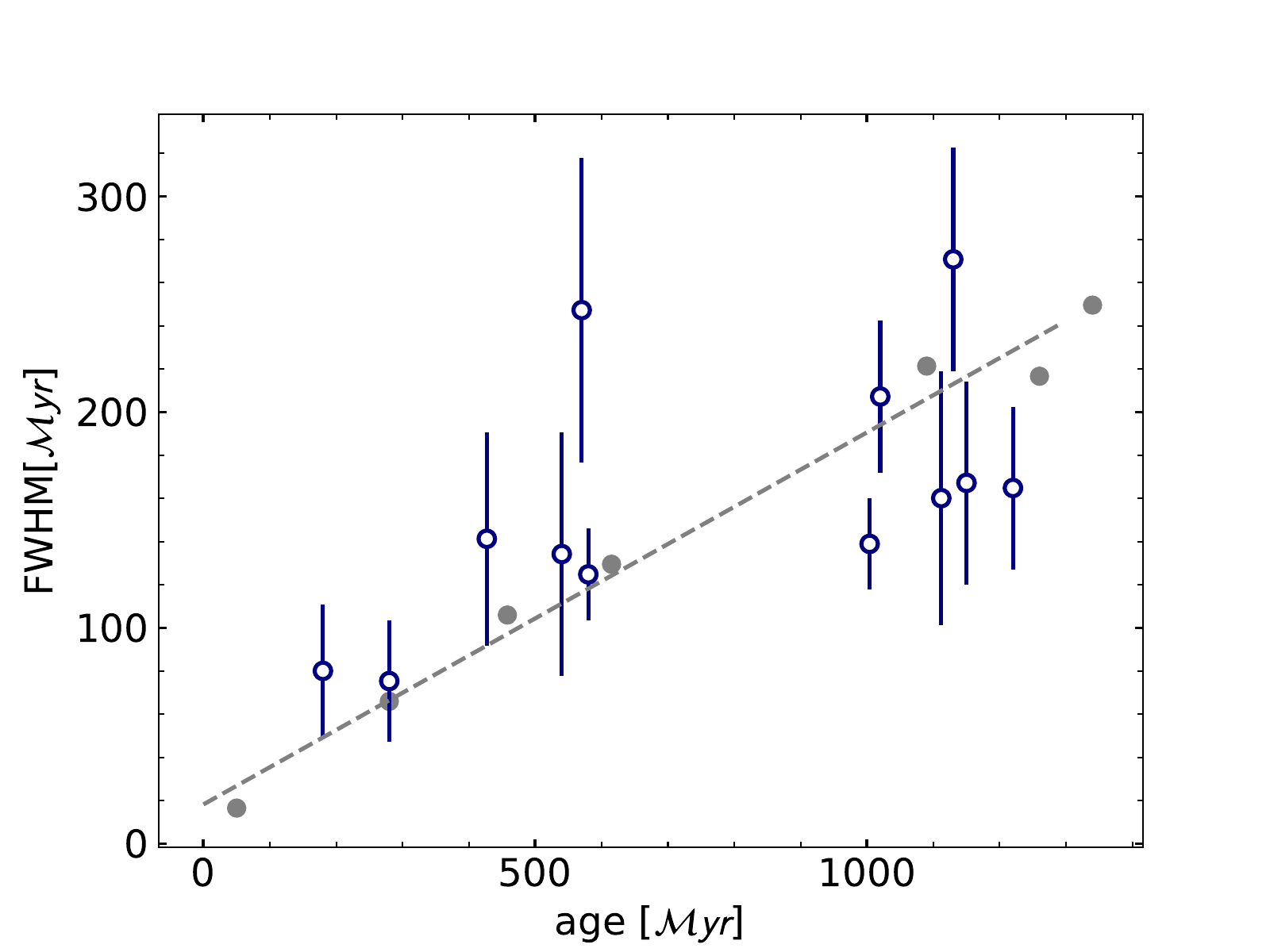}
  \caption{Full width half maximum of the age distribution as a function of cluster age. Blue dots with error bars refer to the analyzed clusters.
     Grey dots are derived from synthetic CMDs of coeval stellar population with different rotation rates. The dashed line is the least-squares best-fit straight line for the gray dots. See text for details.}
 \label{fig:agerel} 
\end{figure}

Our results are summarized in Table~\ref{tab:results} where we provide the full width half maximum of the age distribution, FWHM=$2.355 \cdot \sigma_{\rm AGE}$, for each cluster.
We find that the FWHM ranges from $\sim$70 for NGC\,3114 to $\sim$260~Myr for NGC\,5822 and correlates with the cluster age as shown in Fig.~\ref{fig:agerel}, with old clusters having, on average, larger age spread than younger clusters. 
 A similar trend between the age spread inferred from the eMSTO and the cluster age is also present among Magellanic Cloud clusters and is interpreted as the signature of stellar rotation. 
Indeed, since rotating stars have longer MS lifetime than non-rotating stars with the same age and mass, they would appear younger than coeval non-rotating stars within the same cluster. In this case, if the resulting eMSTO is interpreted as an age spread, the resulting age spread would correlate with the cluster age (Niederhofer et al.\,2015; Bastian et al.\,2018).  
On the other hand, in the case of a true age spread we would expect that the amount of age spread does not depend on cluster age, and therefore a correlation would be very unlikely.

To further investigate the effect of rotation on the observed CMDs, we
extended the method by Niederhofer and collaborators to Galactic open
clusters, 
 and compared the observations with simulated CMDs of coeval stellar populations with different rotation rates based on stellar models from the Geneva database with Z=0.014 and various ages (Mowlavi et al.\,2012; Ekstr{\"o}m et al.\,2013; Georgy et al.\,2014). 
To simulate the CMDs we first retrieved the synthetic photometry corresponding to the best-fit non-rotating isochrones, and for the isochrones with rotation equal to 0.9 times the breakout value ($\omega=0.9\omega_{\rm cr}$). These data account for the limb-darkening effect as in Claret (2000), adopt the gravity-darkening model by Espinosa Lara \& Rieutord (2011) and assume random distribution for the viewing angle.
 We transformed the synthetic photometry into the observational plane
 by adopting the model atmospheres by Castelli \& Kurucz\,(2003) and
 the transmission curves of the $G_{\rm BP}$ and $G_{\rm RP}$ filters
 of Gaia. We assumed that one third of stars in the simulated CMD do
 not rotate, while two thirds of stars have $\omega=0.9\omega_{\rm
   cr}$, in close analogy with what is observed in Magellanic Clouds
 open clusters (e.g.\,Milone et al.\,2018). 

We first applied the procedure above to each synthetic CMD, by
assuming that the eMSTO is due to age spread, and derived the FWHM of
the age distribution. Results are represented with grey dots in
Fig.~\ref{fig:agerel}. As expected, the age spread increases with the
cluster age, in close analogy with what was previously found by
Niederhofer et al.\,(2015) in Magellanic Clouds clusters. The fact that the FWHM values derived for synthetic CMDs and for Galactic open clusters follow similar trends against the cluster age suggests that rotation is the main responsible for the observed eMSTOs.

\begin{figure*} 
  \centering
  \includegraphics[width=8.8cm,height=4.5cm,trim={0cm 0.8cm 0cm 0cm},clip]{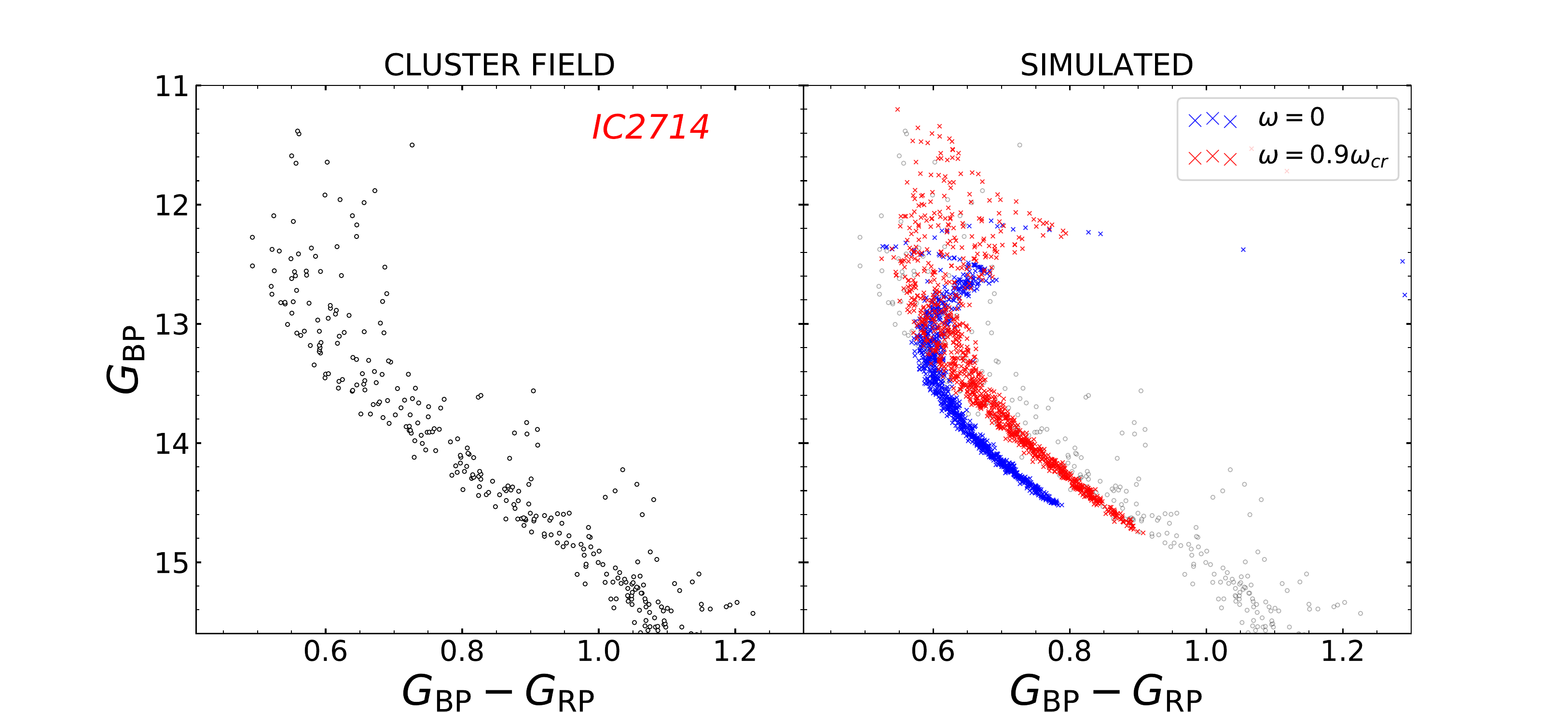}
  \includegraphics[width=8.8cm,height=4.5cm,trim={0cm 0.8cm 0cm 0cm},clip]{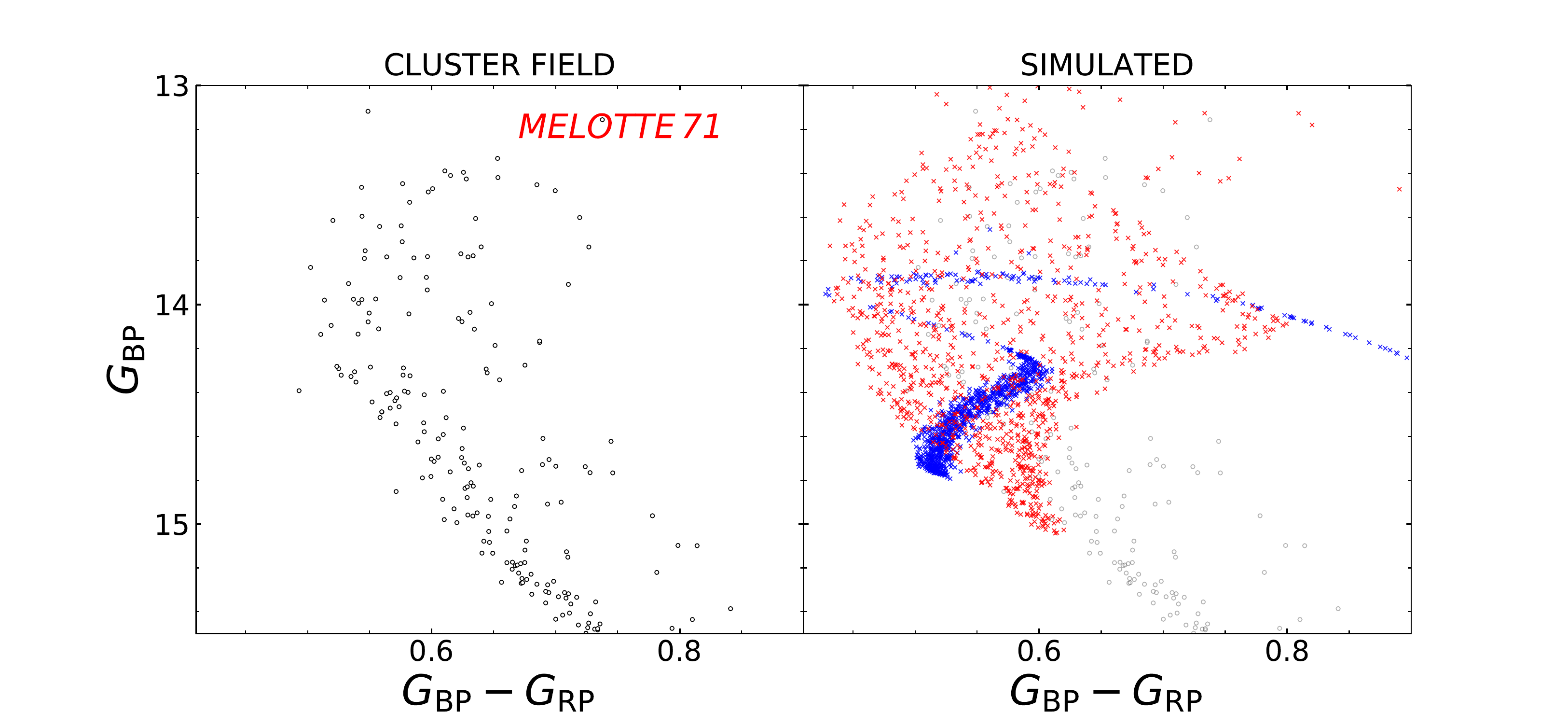}
  \includegraphics[width=8.8cm,height=4.5cm,trim={0cm 0.8cm 0cm 1.6cm},clip]{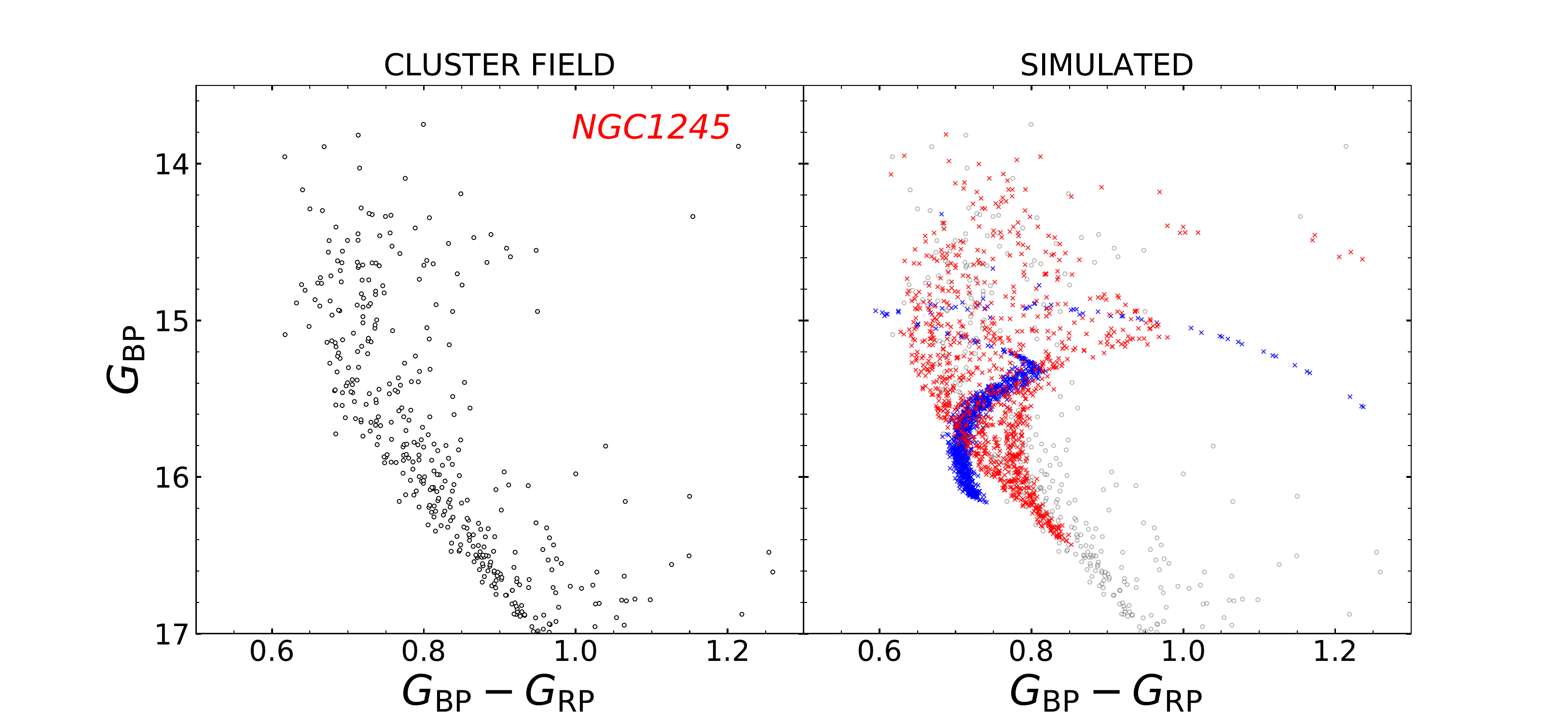}
  \includegraphics[width=8.8cm,height=4.5cm,trim={0cm 0.8cm 0cm 1.6cm},clip]{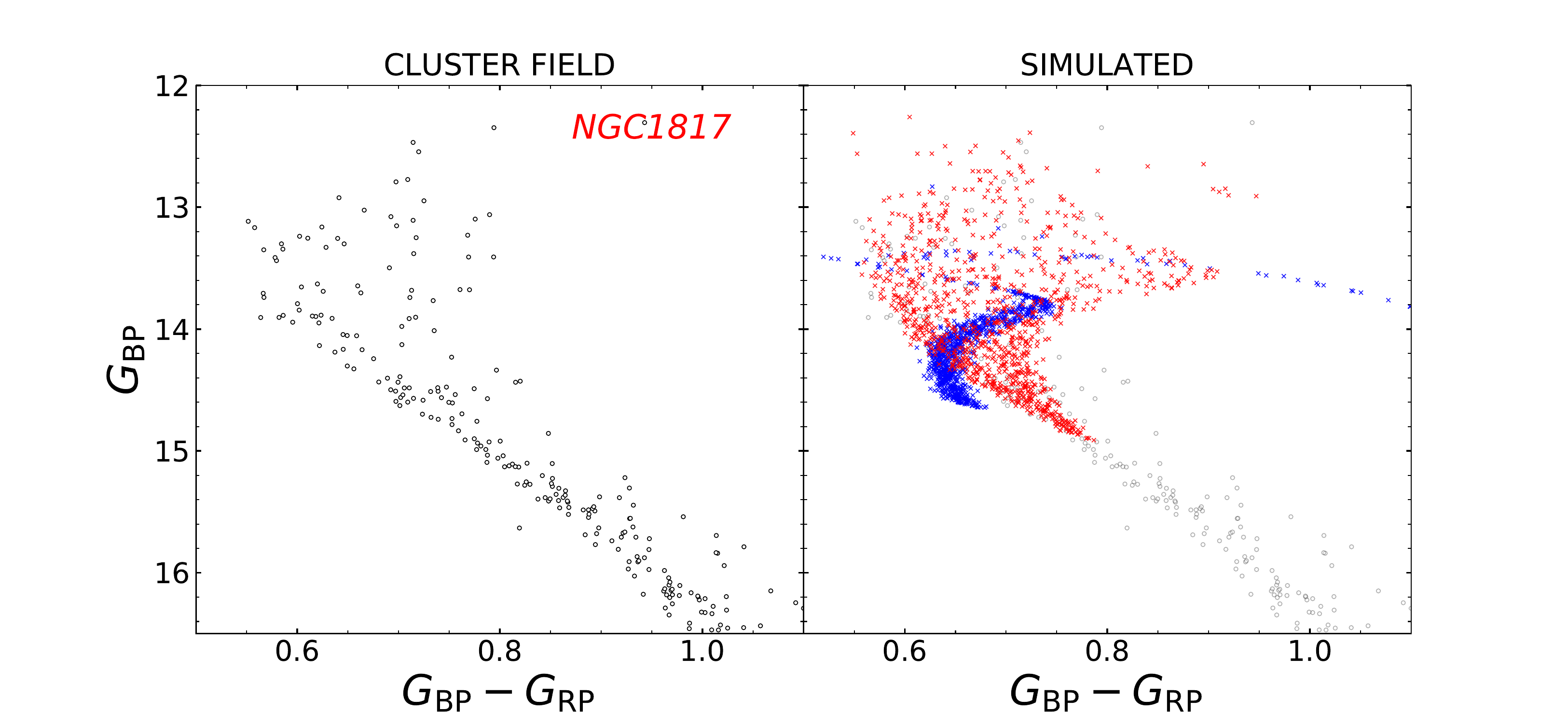}
  \includegraphics[width=8.8cm,height=4.5cm,trim={0cm 0cm 0cm 1.6cm},clip]{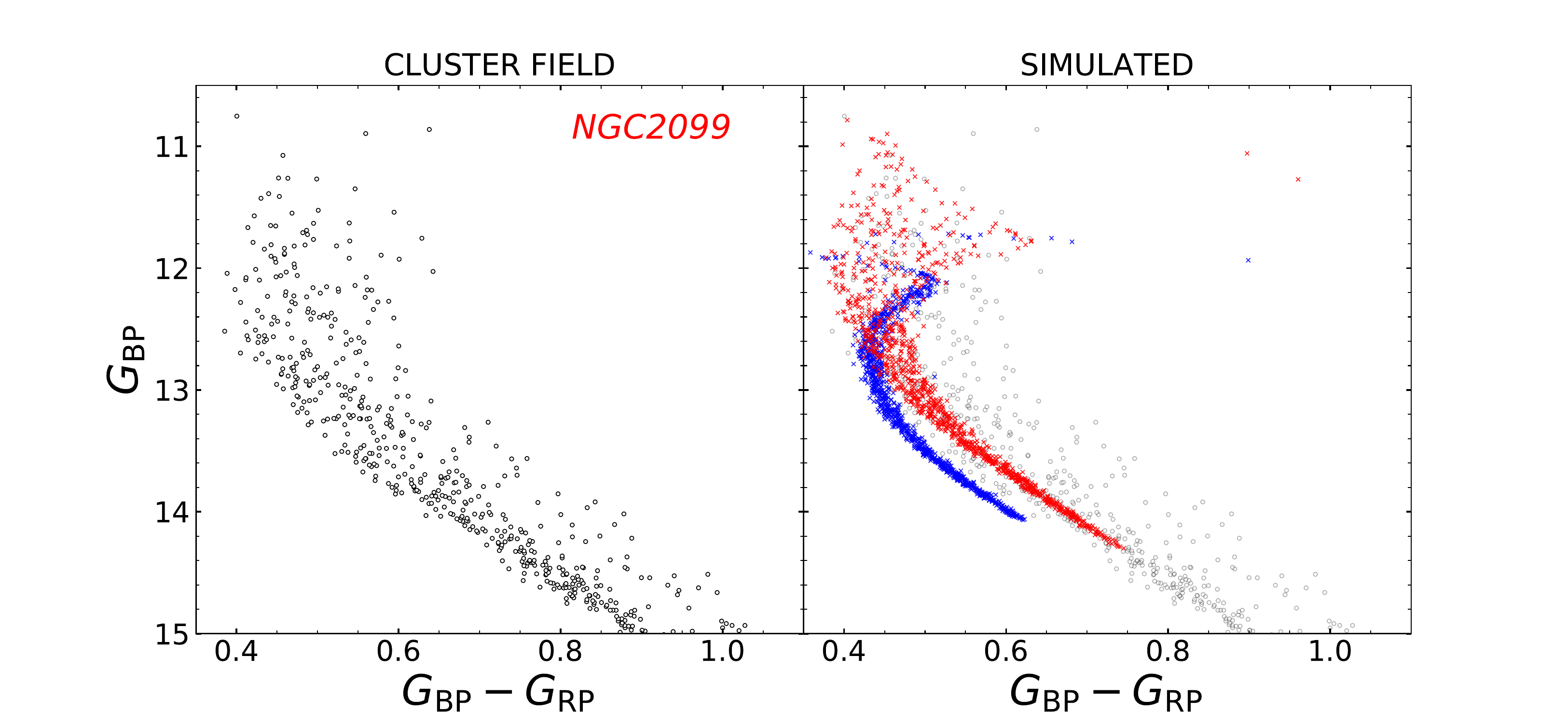}
  \includegraphics[width=8.8cm,height=4.5cm,trim={0cm 0cm 0cm 1.6cm},clip]{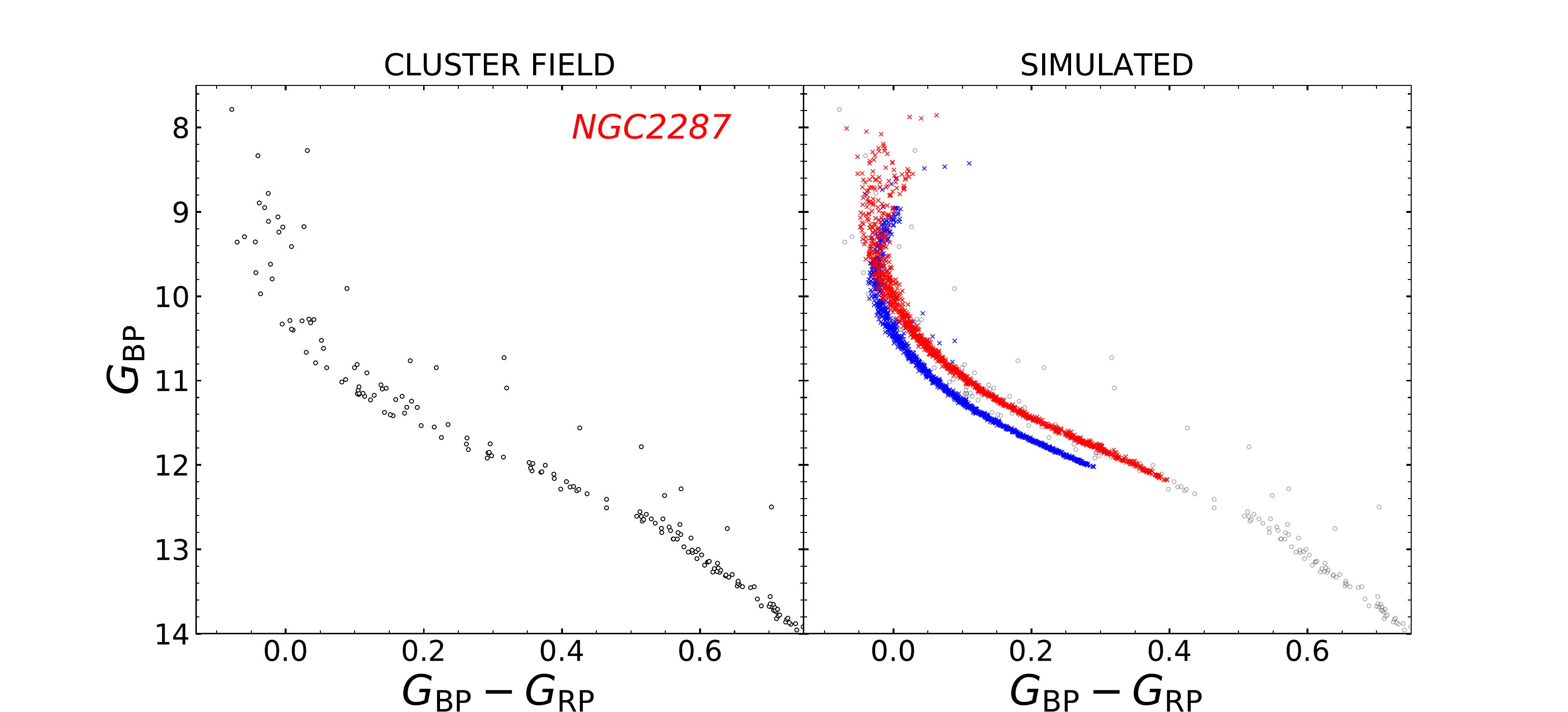}
  \caption{\textit{Left panels.} reproduction of the observed CMDs of cluster members of IC\,2714, Melotte\,71, NGC\,1245, NGC\,1817, NGC\,2099 and NGC\,2287. \textit{Right panels.} Comparison of the observed CMDs plotted in the left panels (grey dots) and simulated CMDs of a non rotating stellar population (blue) and of a stellar population with rotation $\omega=0.9 \omega_{\rm cr}$ (red).}
 \label{fig:rot1} 
\end{figure*}

\begin{figure*} 
  \centering
  \includegraphics[width=8.8cm,height=4.5cm,trim={0cm 0.8cm 0cm 0cm},clip]{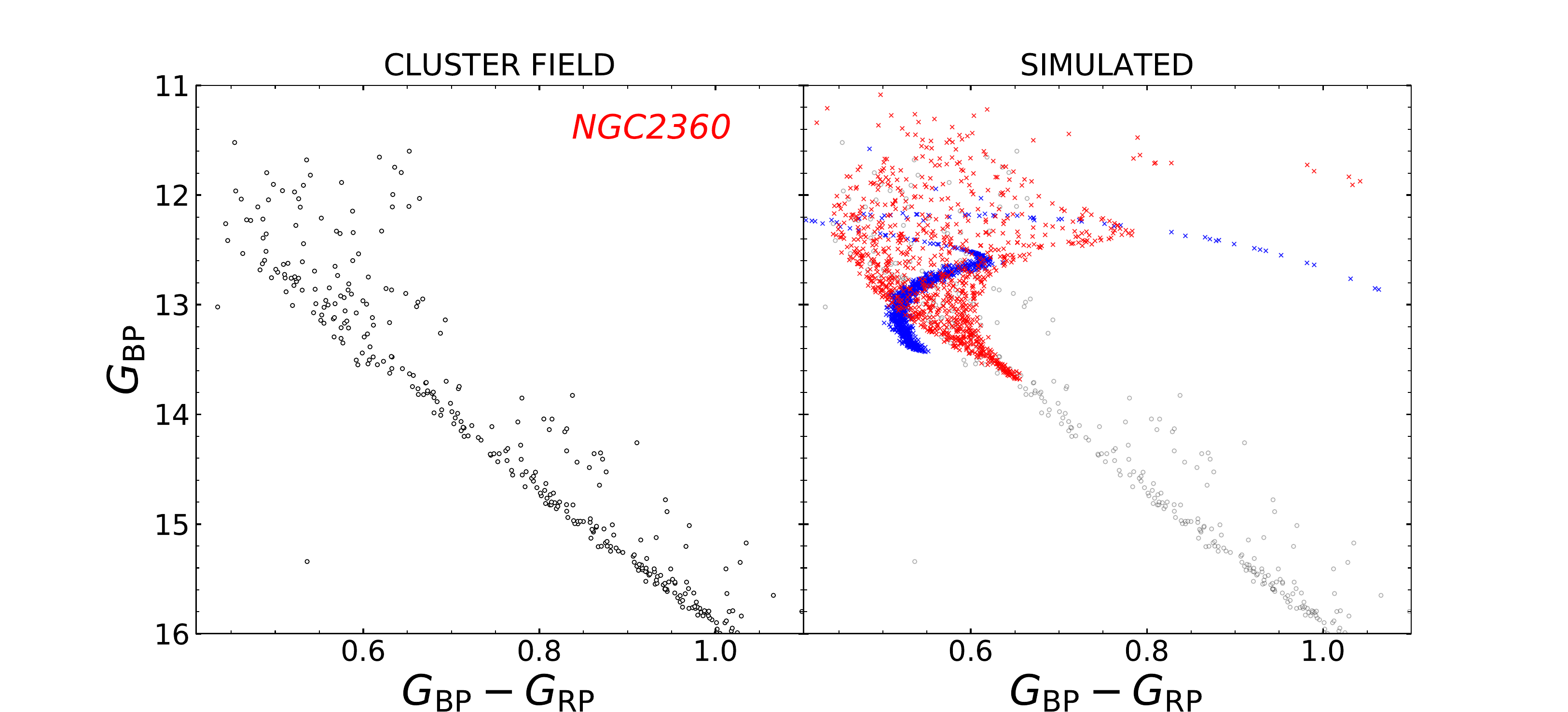}
  \includegraphics[width=8.8cm,height=4.5cm,trim={0cm 0.8cm 0cm 0cm},clip]{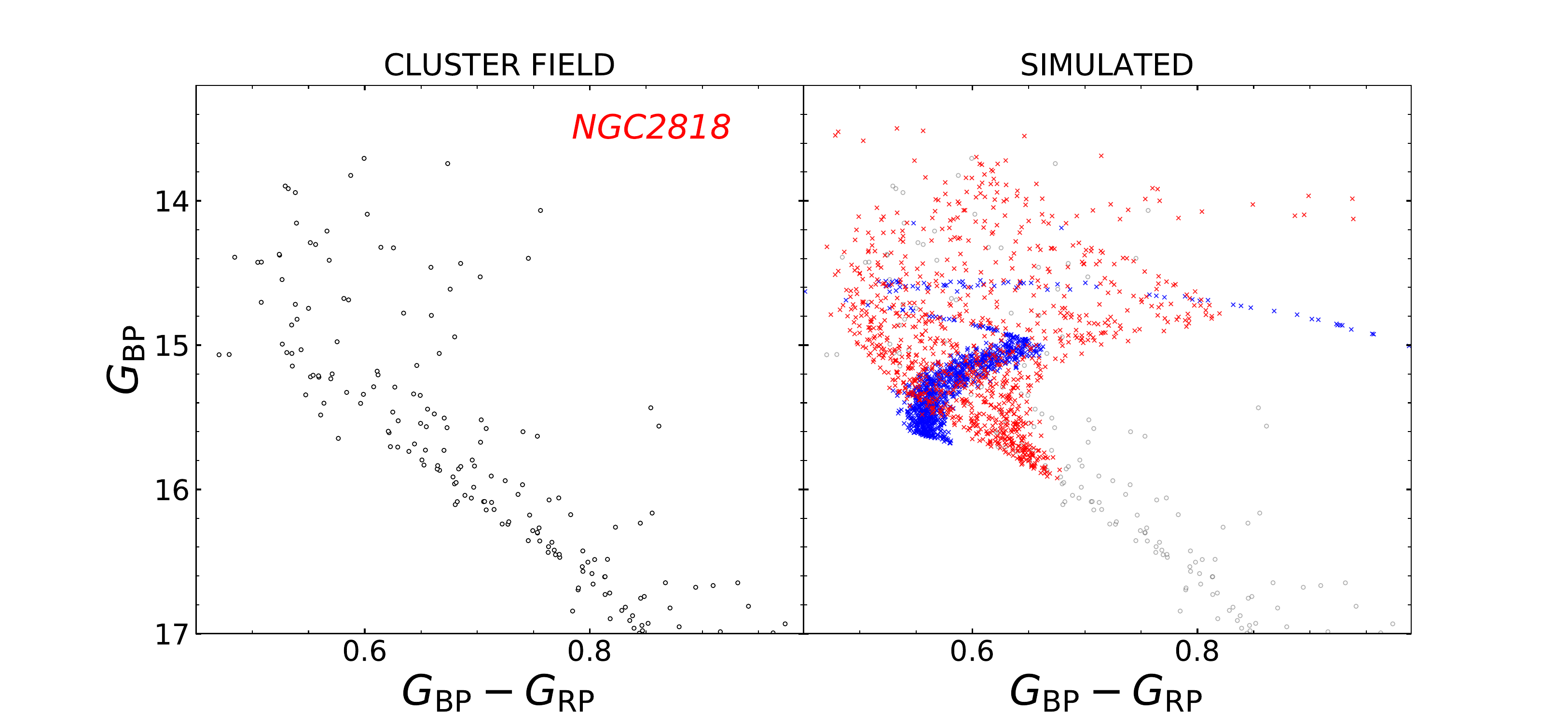}
  \includegraphics[width=8.8cm,height=4.5cm,trim={0cm 0.8cm 0cm 1.6cm},clip]{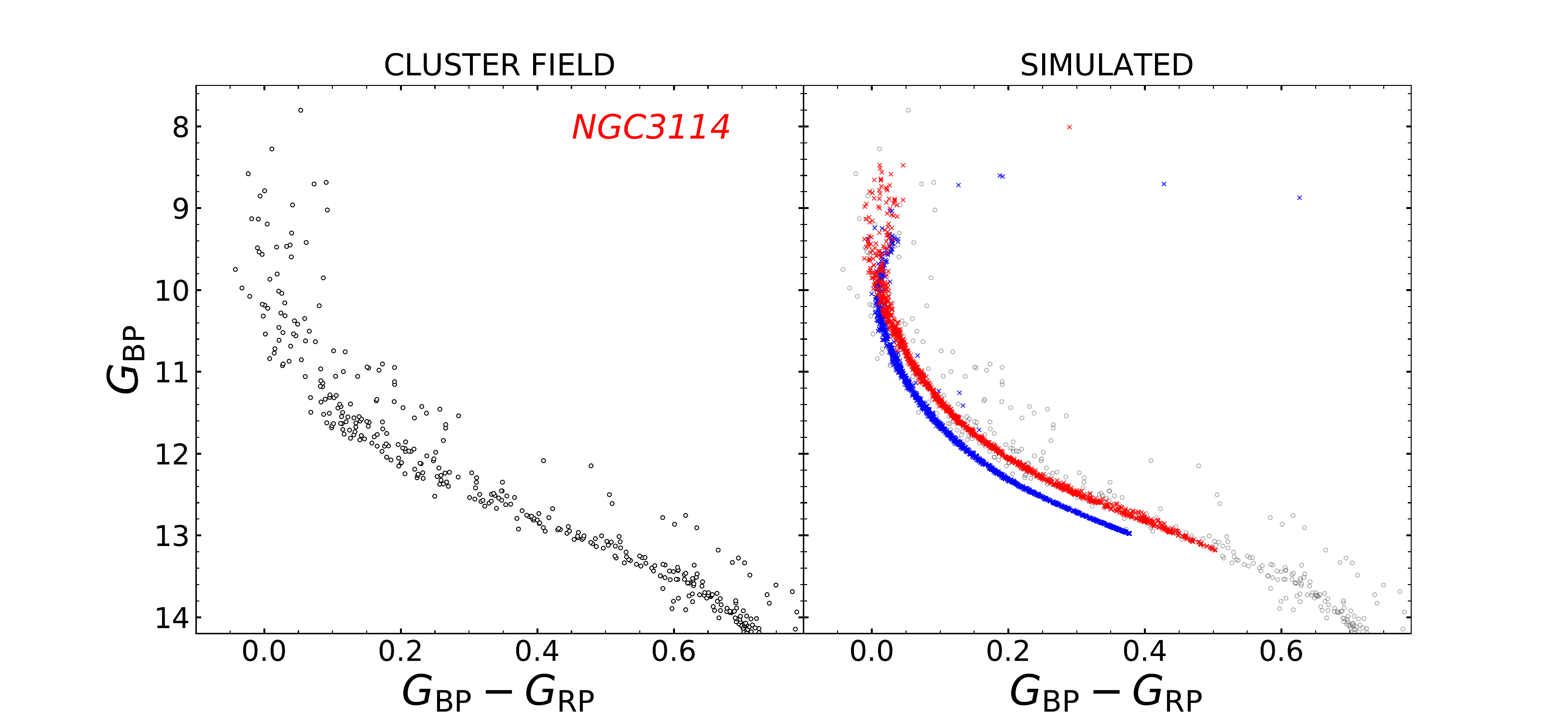}
  \includegraphics[width=8.8cm,height=4.5cm,trim={0cm 0.8cm 0cm 1.6cm},clip]{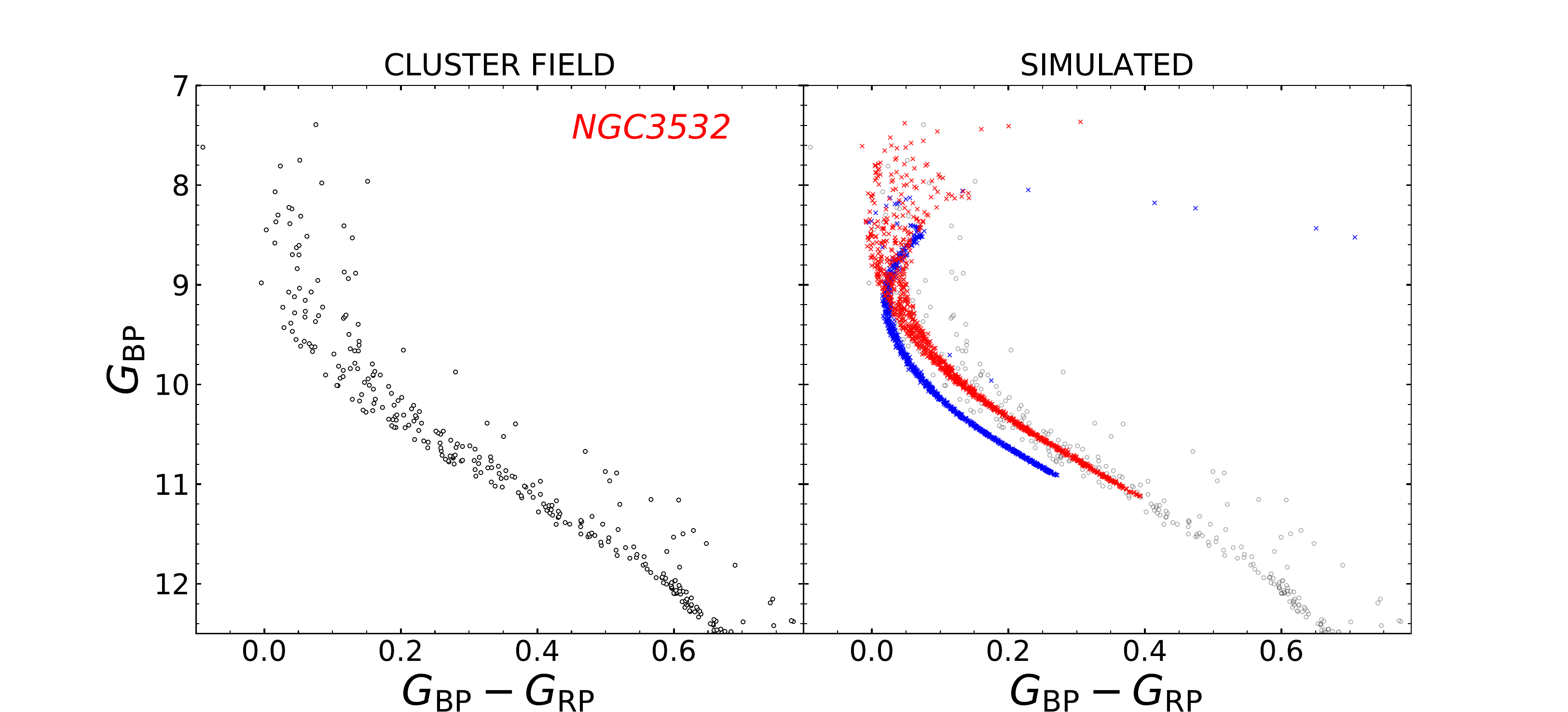}
  \includegraphics[width=8.8cm,height=4.5cm,trim={0cm 0cm 0cm 1.6cm},clip]{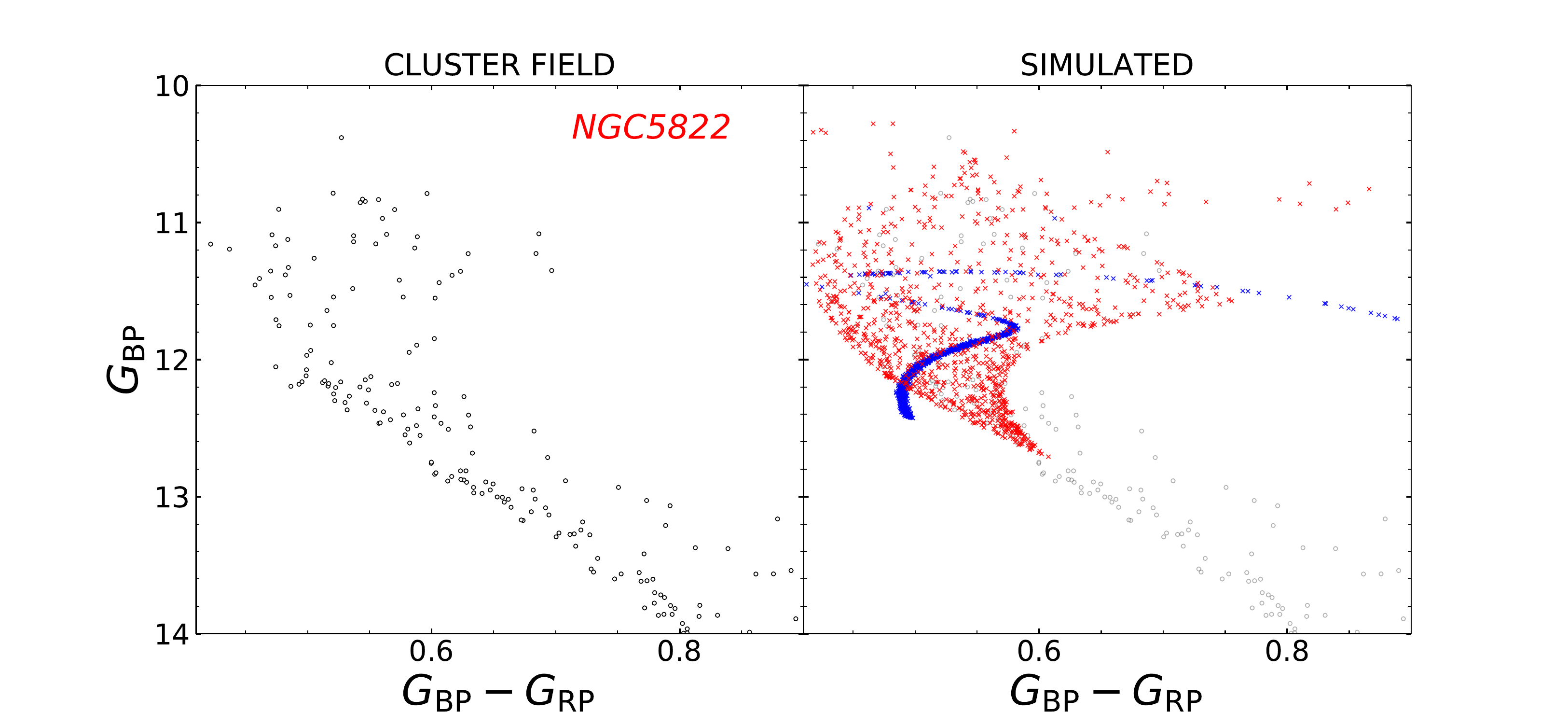}
  \includegraphics[width=8.8cm,height=4.5cm,trim={0cm 0cm 0cm 1.6cm},clip]{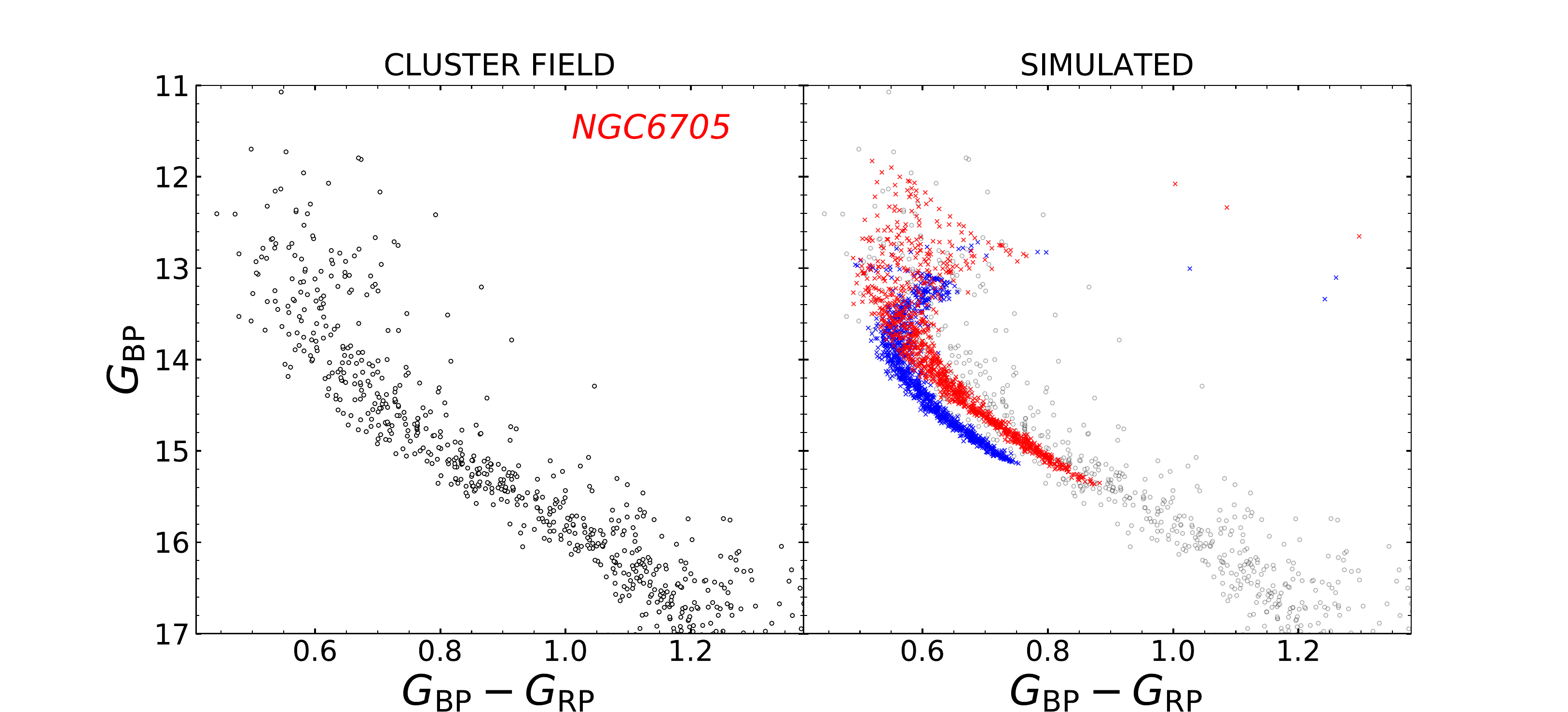}
  \caption{As in Fig ~\ref{fig:rot1} but for NGC\,2360, NGC\,2818, NGC\,3114, NGC\,3532, NGC\,5822 and NGC\,6705}
 \label{fig:rot2} 
\end{figure*}

Finally, we compare in Fig.~\ref{fig:rot1}-~\ref{fig:rot2} the CMDs of cluster members (left panels) with simulated CMDs (right panels). Synthetic CMDs are derived from the Geneva database (Georgy et al.\,2014) and have metallicity, Z=0.014, and similar age, distance modulus and reddening as those listed in Table~1. Unfortunately, rotating models are not available for stars less massive than $\sim$1.7 $\mathcal{M}_{\odot}$.
 e note that, while in young clusters like NGC\,2287 and NGC\,2099 both fast rotators and slow-rotator stars are needed to reproduce the broad MS, the eMSTO of old clusters seems consistent with fast rotators alone.
 The poor quality of the fit could be due to the modeling of several second-order parameters that characterize the end of the core hydrogen burning phase, including the parametrization of the inclination angle, which strongly affects the stellar luminosity and effective temperature (see D’Antona et al.\,2015 for details). Nevertheless, the comparison between data and simulations corroborates the conclusion that stellar rotation is the main responsible for the observed eMSTOs and the broadened MSs. 

\section{Summary and discussion}\label{sec:summary}

We have presented the first analysis of twelve open clusters in
the Milky Way in the context of multiple stellar populations. 
Our results suggest that the multiple photometric sequences observed by Gaia 
in the CMDs of these nearby objects belong to the same
 phenomenon present in Magellanic Clouds clusters, and
interpreted as due to stellar rotation and/or age spreads.

Since the early discoveries, the eMSTOs have been considered a common
feature of the CMDs of LMC and SMC clusters younger than $\sim$2.5~Gyr
whereas the CMDs of Galactic open clusters were thought to be similar
to simple isochrones. 
This picture has been challenged by the recent findings of eMSTOs in
four Galactic open clusters younger than $\sim$1~Gyr, namely
NGC\,2099, NGC\,2360, NGC\,2818, NGC\,6705 (Marino et al.\,2018;
Bastian et al.\,2018).  

 We exploited the Gaia DR2 to analyze the CMDs of twelve Galactic open
 clusters younger than $\sim$1.5~Gyr. We carefully separated cluster
 stars from field stars by using proper motions and parallaxes from
 Gaia DR2 and corrected the photometry of clusters members for
 differential reddening. 
 We find that all the analyzed clusters show the eMSTO. In addition,
 all the clusters younger than $\sim$700~Myr exhibit a broadened upper
 MS, whereas the bottom of the MS is narrow and well defined. The
 appearance of certain photometric features depending on age, is
 similar to that observed in Magellanic Clouds clusters.
 
 We statistically subtracted field stars with cluster-like proper
 motions and parallaxes  from the CMD of candidate cluster stars thus
 demonstrating that eMSTOs and broadened MSs are not due to residual
 contamination from field stars. 
We calculated for each cluster the synthetic photometry of a simple
population of stars with the same age, metallicity, binary fraction,
and observational errors. The comparison between the observations and
the simulated CMDs reveals that the eMSTOs and the broadened MSs are
due neither to observational uncertainties nor to unresolved
binaries. These facts demonstrate that eMSTOs and broadened MSs are
intrinsic features of the CMDs of the analyzed open clusters. 

To investigate the physical mechanisms that are responsible for the eMSTO we first compared the CMDs of cluster members with isochrones with different ages. The eMSTOs of the analyzed open clusters are consistent with stellar populations with different ages in close analogy with what has been observed in Magellanic Cloud clusters with similar ages.
The FWHM of the age spread ranges from about 70~Myr in the $\sim$150-Myr old cluster NGC\,3114 to $\sim$260~Myr in $\sim 1.1$-Gyr old NGC\,5822. Interestingly, the derived age spread correlates with the cluster age, with old clusters having on average larger age spread than young clusters. 
A similar trend between the FWHM of the age distribution and the cluster age is present among Magellanic-Cloud clusters and is interpreted as an evidence that rotation is the main responsible of the eMSTO. Indeed, in a simple stellar population, fast rotators appear younger than coeval non-rotating stars with the same age. 

We compared the CMDs of cluster members with synthetic diagrams
derived from Geneva models and find that the eMSTOs and the broadened
MSs are consistent with coeval stellar populations with different
rotation rates. These findings suggest that rotation is the main
responsible for the eMSTOs and the broadened MSs observed in Galactic
clusters and corroborate direct spectroscopic evidence that stars with
different rotation rates populate the eMSTOs of NGC\,6705 and
NGC\,2818 (Marino et al.\,2018b; Bastian et al.\,2018) and that the
blue and the red MS of NGC\,6705 are populated by slow rotators and
fast rotators, respectively (Marino et al.\,2018b). 

Our investigation of 12 Galactic open clusters demonstrates that the
eMSTO and the broadened MS are not a peculiarity of Magellanic Cloud
star clusters but are common features of Galactic open
clusters. Coeval stellar populations with different rotation rates are
likely the responsible for the eMSTO and the broadened MS of the
analyzed clusters.

\input{par_table.tex}

\section*{acknowledgments} 
\small
 We thank the anonymous referee for several comments that improved the quality of this manuscript. This work has received funding from the European Research Council (ERC) under the European Union's Horizon 2020 research innovation programme (Grant Agreement ERC-StG 2016, No 716082 'GALFOR', PI: Milone), and the European Union's Horizon 2020 research and innovation programme under the Marie Sk\l odowska-Curie (Grant Agreement No 797100, beneficiary: Marino). APM acknowledges support from MIUR through the the FARE project R164RM93XW ‘SEMPLICE’; AFM has been supported by the Australian Research Council through Discovery Early Career Researcher Award DE160100851.

\bibliographystyle{aa}

\end{document}

%% file: par_table.tex
\begin{table*}
\centering
\begin{tabular}{llllllll}
\toprule
\toprule
Cluster & $(m-M)_0$ & E(B$-$V) & Z & Age [Myr] & FWHM [Myr] & $f_{\rm bin}^{\rm q>0.7} $ & $f_{\rm bin}^{\rm tot}$   \\
\midrule
IC\,2714   & 10.60 & 0.38 & 0.0205 & 540  & $134 \pm 57$ & 0.105 & 0.350 \\
MELOTTE\,71& 11.60 & 0.22 & 0.0095 & 1220 & $165 \pm 38$ & 0.085 & 0.283 \\
NGC\,1245  & 12.45 & 0.29 & 0.0183 & 1000 & $139 \pm 21$ & 0.119 & 0.397 \\
NGC\,1817  & 11.00 & 0.26 & 0.0100 & 1030 & $165 \pm 47$ & 0.083 & 0.277 \\
NGC\,2099  & 10.90 & 0.26 & 0.0300 & 580  & $125 \pm 21$ & 0.085 & 0.283 \\
NGC\,2287  & 9.40  & 0.04 & 0.0219 & 280  & $76  \pm 28$ & 0.034 & 0.113 \\
NGC\,2360  & 10.23 & 0.16 & 0.0140 & 1020 & $210 \pm 35$ & 0.087 & 0.290 \\
NGC\,2818  & 12.35 & 0.22 & 0.0100 & 1110 & $160 \pm 59$ & 0.088 & 0.293 \\
NGC\,3114  & 10.05 & 0.12 & 0.0209 & 180  & $75  \pm 31$ & 0.068 & 0.227 \\
NGC\,3532  & 8.30  & 0.06 & 0.0160 & 430  & $140 \pm 50$ & 0.074 & 0.247 \\
NGC\,5822  & 9.40  & 0.11 & 0.0170 & 1130 & $270 \pm 52$ & 0.131 & 0.437 \\
NGC\,6705  & 11.10 & 0.46 & 0.0083 & 570  & $245 \pm 71$ & 0.153 & 0.510 \\
\bottomrule
\bottomrule
\end{tabular}
\caption{Distance modulus, reddening, age, FWHM of the age distribution, fraction of binaries with mass-ratio, q$>$0.7, and total fraction of binaries inferred in this paper. Cluster metallicities are from Paunzen et al.\,(2010).}
\label{tab:results}
\end{table*}